\documentstyle[epsf,aps]{revtex}
\renewcommand{\thefootnote}{\fnsymbol{footnote}}
\begin{document}
\newcommand{\be}{\begin{eqnarray}}
\newcommand{\dlq}{\lq\lq}
\newcommand{\ee}{\end{eqnarray}}
\newcommand{\ben}{\begin{eqnarray*}}
\newcommand{\een}{\end{eqnarray*}}
\newcommand{\stackeven}[2]{{{}_{\displaystyle{#1}}\atop\displaystyle{#2}}}
\newcommand{\lsim}{\stackeven{<}{\sim}}
\newcommand{\gsim}{\stackeven{>}{\sim}}
\newcommand{\un}[1]{\underline{#1}}
\renewcommand{\baselinestretch}{1.0}
\newcommand{\as}{\alpha_s}
\newcommand{\tas}{{\tilde\alpha}_s}
\def\eq#1{{Eq.~(\ref{#1})}}
\def\fig#1{{Fig.~\ref{#1}}}
\begin{flushright}
NT@UW--03--012 \\
BNL-NT-03/9 
\end{flushright}
\vspace*{1cm} 
\setcounter{footnote}{1}
\begin{center}
{\Large\bf Baryon Stopping and Valence Quark Distribution at Small $x$}
\\[1cm]
Kazunori Itakura$^{1}$, Yuri V.\ Kovchegov$^{2}$, Larry McLerran$^{3}$
and Derek Teaney$^{3}$ \\ ~~ \\ 
{\it $^1$ RIKEN BNL Research Center,
Brookhaven National Laboratory,  Bldg. 510A} \\ {\it Upton NY, 11973} \\ ~~ \\ 
{\it $^2$ Department of Physics, University of Washington, Box 351560} \\
{\it Seattle, WA 98195 } \\ ~~ \\ 
{\it $^3$ Nuclear Theory Group, Brookhaven National Laboratory, Bldg. 510A} \\ 
{\it Upton, NY 11973} \\ ~~ \\ ~~ \\
\end{center}
\begin{abstract}
We argue that the amount of baryon stopping observed in the central
rapidity region of heavy ion collisions at RHIC is proportional to the
nuclear valence quark distributions at small $x$. By generalizing
Mueller's dipole model to describe Reggeons we construct a non-linear
evolution equation for the valence quark distributions at small $x$ in
the leading double-logarithmic approximation. The equation includes
the effects of gluon saturation in it.  The solution of the evolution
equation gives a valence quark distribution function $dn_{val}/dy \sim
e^{-(0.4\div0.5) \, y}$. We show that this $y$-dependence as 
well as the predictions of Regge theory are consistent with
the net-proton rapidity distribution reported by BRAHMS.
\end{abstract}

\renewcommand{\thefootnote}{\arabic{footnote}}
\setcounter{footnote}{0}

\section{Introduction}

\begin{figure}   
\begin{center} 
\epsfxsize=7.5cm
\leavevmode
\hbox{ \epsffile{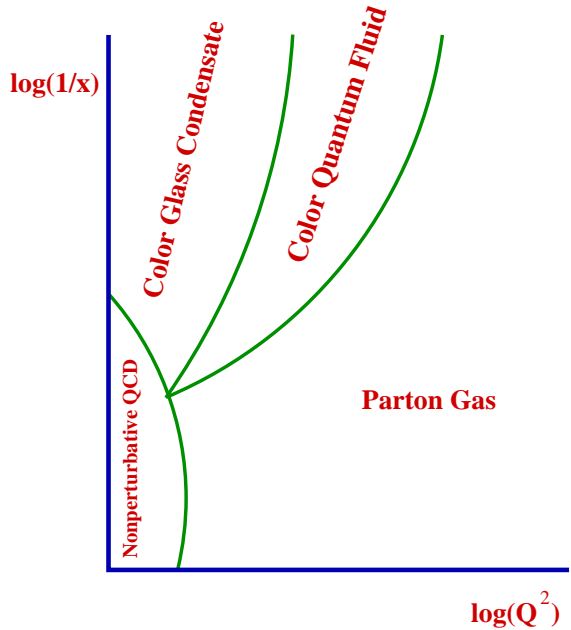}}
\end{center}
\caption{The various phases of high density QCD. (For specific numerical
estimates see \protect\cite{LL}). In the Color
Glass Condensate the gluon density is large.  In the Color Quantum
Fluid phase, the density is low, but correlation functions are power
law behaved with anomalous dimensions.  In the parton gas phase, the
density is described by the ordinary evolution equations of DGLAP or
BFKL equation. }
\label{phase}
\end{figure}

One of the outstanding theoretical issues related to our understanding
of hadrons is the problem of how distributions of quarks and gluons
arise.  In the last several years, remarkable progress has been made
in our understanding of the gluon distribution at small $x$
\cite{glrmq,mv,kjklw,yuri,dip,bal,JKLW,FILM,IM}. 
One has been able to show the 
existence of a region of a high density of gluons in a very coherent
configuration, the Color Glass Condensate and to understand how this
region joins on to the low density region
\cite{mv,kjklw,JKLW,FILM,LT,IIM,MT,Gaussian}. These regions correspond 
to different values of $x$ and $Q^2$ for which the gluon distribution
function is measured.  The region of low density can be understood by
a combination of techniques associated with the BFKL \cite{BFKL} and
the DGLAP \cite{dglap} evolution equations.  Within the low density
region, it has been found that the geometric scaling \cite{Stasto}
extends outside of the saturation region, where the correlation
functions for gluons are pure powers with anomalous dimensions
\cite{IIM}.  At larger $Q^2$ and or larger values of $x$, the density
of gluons is even lower and results match on smoothly to the results
of DGLAP evolution \cite{dglap}.  These various regions are shown in
\fig{phase}.  (For specific numerical estimates see \cite{LL}.)

The essential feature of small $x$ physics which allows for a solution
for the gluon distribution is that the density of gluons is very large
\cite{mv}.  This density can be thought of as a momentum scale
squared, $Q_s^2$, where the subscript $s$ refers to saturation
\cite{glrmq,mugla}.  Saturation here means that the density of gluons
at any given value of $x$ approaches a fixed limit as we go toward
smaller $x$.  For $p_T^2 \le Q_s^2$, the gluon phase space density
in a hadron or nucleus of radius $R$ is \cite{mv}
\be
	{1 \over {\pi R^2} } {{dn_G} \over {dy \, d^2p_T}} \sim {1 \over
	\as}.
\ee
Because the strong coupling constant is evaluated at $Q_s^2 \gg
\Lambda^2_{QCD}$, the coupling is small.  The phase space density is large,
and hence the gluons are in a condensate.  The glass arises because
the gluons are described by classical fields which are produced by
sources at higher values of $x$ and therefore have their time scales
Lorentz time dilated. Furthermore, the gluon distribution function is
somewhat disordered in the transverse plane \cite{JKLW,FILM}
suggesting glassy behavior. Since the coupling constant is weak, weak
coupling methods may be combined with renormalization group techniques
to solve for the properties of the Color Glass Condensate
\cite{yuri,bal,JKLW,FILM}.  Because in the weak field region, the
typical momentum scale is large, there is no infrared problem in
solving BFKL or DGLAP evolution.  The infrared cutoff is in fact the
saturation momentum as this is the scale on which the gluon color
charge distribution neutralizes
\cite{mv,kjklw,IM,Gaussian,Al-neutral}.

A problem currently not understood is the origin of the valence quark
number distributions at small $x$.  This includes both the baryon
number distribution and isospin number distribution.  Spin is also a
valence effect for the distributions of quarks and gluons inside a
hadron, but spin, unlike baryon number and isospin, can be carried by
the gluons.  We shall restrict our attention here to isospin and
baryon number since it is a little simpler, although many of the
techniques developed here might be applied to this case.  In the case
of baryon number, it is true that non-perturbative topological
excitations of the gluon field might give contributions to baryon
number \cite{Veneziano,KZ,Dima}.  For weak coupling, we expect these
contributions to be small, and in this paper we shall not consider
their effect.  It would be most interesting to have a proper first
principle computation of the magnitude of these effects within the
Color Glass Condensate.  An experimental analysis of the stopping
of net electric charge versus the stopping of baryon number, which
would allow one to disentangle the contributions of the valence quarks
from the non-perturbative topological gluonic excitations at RHIC is
under way \cite{Stankus}.

We shall assume here that baryon number and isospin are carried by
valence quarks \cite{EK,BMS}.  In weak coupling, we should be able to
compute the $x$ and $Q^2$ dependence of these valence quark
distribution functions.  We will include non-perturbative aspects of
the Color Glass Condensate in our computation since the quarks,
although weakly coupled, propagate in the strong background field
associated with the Color Glass Condensate.

In experiment, one typically measures the valence quark distribution
in the distribution of particles produced in the final state.  In this
paper, we shall only consider how this is affected by the valence
quark distribution associated with the hadron's wavefunction.  We
shall not consider the effects of final state scattering.  In so far
as this final state scattering is due to local particle interactions,
valence quantum numbers should spread diffusively, and cannot disperse
over a wide range in $x$.  On the other hand, novel particle
interactions have been proposed, often involving the interpretation of
baryon number as a topological excitation of gluons, which allow for
transfer of baryon number \cite{KZ,Dima}. Computation of such effects
is beyond the scope of this paper, but they might be possible to
include in a computation such as that advocated by Krasnitz and
Venugopalan \cite{kv} and by Lappi \cite{Lappi}.

In this paper, we will derive the $x$ and $p_T$ dependence of the
unintegrated valence quark distribution functions, subject to the
caveats above.  To understand the essential role that the Color Glass
Condensate plays in this computation, consider the diagrams of 
Fig.~\ref{ladder} 
computed by Kirschner and Lipatov \cite{KL,Kirschner1,Kirschner2} and by
Griffiths and Ross \cite{GR}:
\begin{figure}   
\begin{center} 
\epsfxsize=8cm
\leavevmode
\hbox{ \epsffile{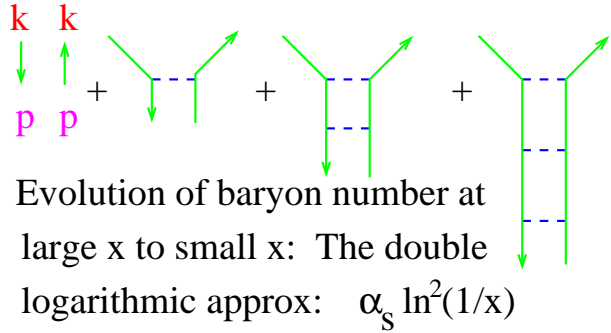}}
\end{center}
\caption{The ladder diagrams for the evolution of baryon number.}
\label{ladder}
\end{figure}
\noindent These diagrams are singular, and the expansion parameter for
individual diagrams is $\alpha_s \ln^2(1/x)$.  The $\ln^2(1/x)$ arises
from the longitudinal and transverse phase space due to the fact that
the $p_T$ integral is not limited in the ultraviolet and is
effectively cut off by the center of mass energy of the system
\cite{KL,Kirschner1,Kirschner2,GR}.

For gluons, the expansion parameter for ladder diagrams is $\alpha_s
\ln(1/x)$, and the result of summation is that \cite{BFKL}
\be
      {{dn_{G}} \over {dx}} \sim {1 \over {x^{1+C\alpha_s}}}.
\ee
For the case of valence quark distributions, we expect that the double
logarithmic behavior is generated by $1/x^{\sqrt{B\alpha_s}}$, so it
is not too surprising that the result of computing diagrams in
Fig. \ref{ladder} is \cite{KL}
\be
	{{dn_{val}} \over {dx}} \sim {1 \over {x^{\sqrt{2 \alpha_s C_F
	/\pi}}}}.
\ee
If we take $\alpha_s \sim 0.25 - 0.5$, we find $dn_{val} / dx \sim
1/x^{0.5 - 0.7}$. Such a value is typical of the Color Glass
Condensate and the valence quark distribution, as we shall see in a
later section, correctly describes the baryon number distribution seen
at RHIC.

The problem with summing the ladder graphs is justifying a weak
coupling expansion \cite{Bartels}.  If the coupling is weak, then such
diagrams indeed generate the leading order contribution.  On the other
hand if one takes the ladder contribution seriously, the dominant
contribution for baryon number occurs at a transverse momentum scale
of order $\Lambda_{QCD}$, and the weak coupling methods fail.  We will
see that properly including the effects of the Color Glass Condensate
generate a cutoff at $p_T \sim Q_s \gg \Lambda_{QCD}$ at
asymptotically small $x$.

To see how this works, consider the diagrams of Fig. \ref{ladder1}:
\begin{figure}   
\begin{center} 
\epsfxsize=8cm
\leavevmode
\hbox{ \epsffile{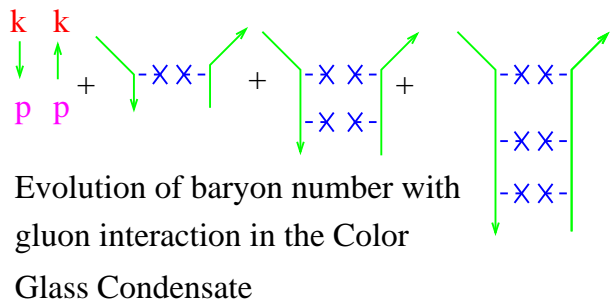}}
\end{center}
\caption{The ladder diagrams for the evolution of baryon number
including the possibility of interaction with the Color Glass
Condensate.}
\label{ladder1}
\end{figure}
\noindent In this figure, one has a ladder sum, but the gluons can be 
absorbed in the Color Glass Condensate. In terms of traditional
Feynman diagrams this implies that any number of gluonic ladder fan
diagrams \cite{glrmq,yuri} can connect to the reggeon (quark) ladder
depicted in \fig{ladder1}. The overall interaction would look like set
of fan diagrams consisting of a single quark ladder with any number of
gluonic (BFKL \cite{BFKL}) ladders attached to it and to each
other. The fan diagrams for gluons have been summed before in
\cite{glrmq,yuri,bal}. Here we are going to write down an equation
summing up the fan diagrams with a single quark ladder. For small
enough $k_T$ of the gluons in \fig{ladder1}, the gluons do not
propagate, they get absorbed in the target and the distribution in
$p_T$ of the quarks does not evolve.  Our expectation is therefore
that we get a growing distribution function for valence quarks only if
the transverse momentum of the quarks exceed the saturation momentum.
The coupling constant should therefore be evaluated at $Q_s$.

The goal of this paper is to show that this is true by constructing
and solving an evolution equation for the valence quark distribution
functions. The equation is constructed in \eq{eqr2} and its solution
with a simple model for gluon evolution is given by \eq{sol}.

\begin{figure}   
\begin{center} 
\epsfxsize=10cm
\leavevmode
\hbox{ \epsffile{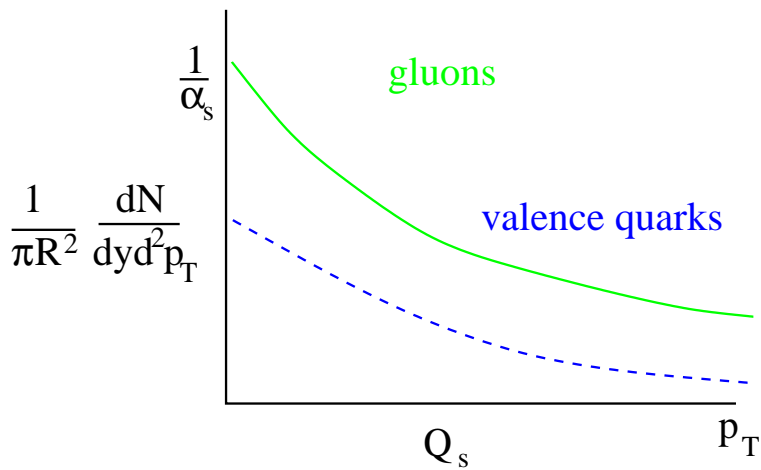}}
\end{center}
\caption{The phase space distribution of valence quarks
and gluons as a function of $p_T$ at fixed $x$.}
\label{distribution}
\end{figure}

The resulting picture which arises for the gluon and quark
distribution functions is amusing: The gluons have phase space density
which is dominated by glue with $k_T \le Q_s$ and has a logarithmic
divergence $\sim \ln Q_s/k_T$ in the infrared.  The valence quarks
have a similar transverse phase space density, which is lower than the
gluon one by powers of Bjorken $x$ (see Eqs. (\ref{conc1}),
(\ref{eqsum}) and (\ref{conc2}) and is described by the quark
saturation scale $Q_s^{quark} = (2/3) \, Q_s$. This is shown in
Fig. \ref{distribution}.

A remarkable conclusion of this work is that for $p_T$ far outside the
saturation region, the valence quark distributions are power law
behaved with non-integer powers (for fixed coupling).  This reflects
the power law behavior with fractional anomalous dimensions found for
the gluon distribution function in the geometrical scaling region
\cite{LT,IIM}.  It adds weight to the conjecture that there is an
intermediate phase between that of the Color Glass Condensate and the
parton gas, the Color Quantum Fluid \cite{Dima2}.

The outline of this paper is as follows: In Sect.~\ref{MVQUARK} we
construct the analog of the McLerran-Venugopalan model for valence
quarks \cite{mv}. This illustrates with a simple model how the
saturation scale regulates the valence quark distribution in the
infrared. Next, in Sect.~\ref{QUANTUM} we derive a small $x$ evolution
equation for the valence quark distribution within Mueller's dipole
picture. This evolution equation is then solved in the linear regime
in Sect.~\ref{linear} and the $x$-intercept is found to agree with
previous results based upon the summation of ladder diagrams
\cite{KL}. Then in Sect.~\ref{nonlinearSection} we solve the full
non-linear evolution equation using a simple theta function model of
the dipole scattering amplitude. This section therefore merges
Sect.~\ref{MVQUARK} and and Sect.~\ref{linear} illustrating how the
saturation scale serves as an infrared cutoff and how quantum
evolution changes the canonical dimensions in the McLerran-Venugopalan
model. Finally, in Sect.~\ref{dataSection} we analyze the rapidity
distribution of net-protons from the BRAHMS experiment \cite{BRAHMS}
and extract a phenomenological intercept.  This intercept is compared
with the perturbative intercept of Sect.~\ref{linear} and the
intercept expected from Regge theory.

\section{Valence Quark Distribution in the Semi-Classical Approximation}
\label{MVQUARK}

In this Section we construct a soft valence quark wave function of a
nucleus in the quasi-classical approximation of McLerran-Venugopalan
model \cite{mv}. This quasi-classical wave function will have some
qualitative features of the full answer and will also serve as the
initial condition for the evolution equation which we will construct
below.

\begin{figure}
\begin{center}
\epsfxsize=5cm
\leavevmode
\hbox{ \epsffile{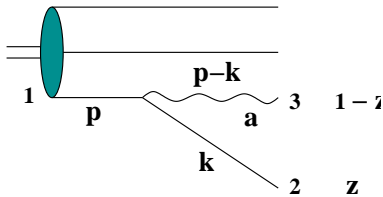}}
\end{center}
\caption{Soft valence quark wave function of a nucleon.}
\label{lo}
\end{figure}

Let us consider an ultrarelativistic nucleus moving in the light cone
``plus'' direction. Similar to \cite{kjklw} we begin by constructing
soft valence quark distribution in a single nucleon at the lowest
order in the coupling as depicted in \fig{lo}. There a valence quark
with momentum $p$ splits into a quark with momentum $k$ and a gluon
with momentum $p-k$. Using the rules of light-cone perturbation theory
\cite{lb} the wave function of the valence quark in $A_+ = 0$ light
cone gauge can be written as (see Appendix \ref{wavefunction})
\be\label{psim}
\psi^a_{\sigma\lambda} ({\un k}, {\un p}- {\un k}, z) \, = \, g \, T^a \, 
[1 + z - \sigma \lambda (1 - z)] \
\frac{{\un \epsilon}^\lambda \cdot ({\un k} - z \, {\un p})}
{({\un k} - z \, {\un p})^2},
\ee
where $z = k_+/p_+$, $\sigma$ is the quark's helicity which is
conserved in the splitting since quarks are assumed to be massless,
$\lambda$ and $a$ are gluon's polarization and color and ${\un
\epsilon}^\lambda$ is the gluon's polarization vector. Transforming into 
transverse coordinate space we end up with
\be\label{psic}
\psi^a_{\sigma\lambda} ({\un x}_{23}, {\un x}_{31}, z) \, = \, g \, T^a \,
[1 + z - \sigma \lambda (1 - z)] \, \delta^2 ({\un x}_{31} + z \, {\un
x}_{23}) \, \frac{i}{2 \pi} \, \frac{{\un \epsilon}^\lambda \cdot
{\un x}_{23}}{{\un x}_{23}^2},
\ee
where ${\un x}_{ij} = {\un x}_i - {\un x}_j$ with ${\un x}_1$ and
${\un x}_2$ the coordinates of the quark before and after the
splitting and ${\un x}_3$ the transverse coordinate of the gluon (see
\fig{lo}).
Squaring the expression in \eq{psic}, summing over gluon
polarizations, averaging over quark helicities and integrating over
the initial quark position ${\un x}_1$ (in the amplitude and in the
complex conjugate amplitude separately) we end up with
\be\label{split}
\Psi (z) \, = \, \int \frac{d^2 x_{23}}{2 \pi} \int_{z_i}^1 \frac{d z}
{2 (1-z)}
\, \frac{1}{2} \, \sum_{\sigma, \lambda, a} |\psi^a_{\sigma\lambda} 
({\un x}_{23},z)|^2 \, = \,  
\int d^2 x_{23} \, \int_{z_i}^1 \, d z \, \frac{{\tilde \alpha}_s}{2 \pi} 
\, \frac{1+z^2}{1-z} \, \frac{1}{x_{23}^2}, 
\ee
where we defined
\be
{\tilde \alpha}_s \, = \, \frac{\as C_F}{\pi},
\ee
$z_i$ is some initial light cone momentum fraction and $C_F =
\frac{N_c^2 -1}{2 \, N_c}$. Looking at the $z$-dependence of
\eq{split} we recognize the real part of the DGLAP splitting function
$\gamma_{qq}$ \cite{dglap}. By transforming $z
\rightarrow 1-z$ we would obtain $\gamma_{Gq}$ splitting function as
expected.

Until now we have not imposed any restrictions on the quark's
longitudinal momentum fraction $z$. Imposing $z \ll 1$ for soft quarks
we get ${\un x}_1 = {\un x}_3$ from \eq{psic}. The soft valence quark
wave function can be written as
\be\label{psis}
\psi^a_{\sigma\lambda} ({\un x}_{21}, z) \, = \, g \, T^a \,
(1 - \sigma \lambda) \, \frac{i}{2 \pi} \, \frac{{\un \epsilon}^\lambda \cdot
{\un x}_{21}}{{\un x}_{21}^2}. 
\ee
Let us define a distribution function of valence quarks in a nucleus
similar to how it was done for gluons in \cite{kjklw,KM}. To do that
we have to multiply the wave function in \eq{psis} by its complex
conjugate at a different transverse coordinate of the soft quark,
average over helicities, sum over polarizations and integrate over
transverse coordinate positions of the initial quark ${\un x}_1$. The
result yields
\be\label{vqd1}
\frac{d n_{val} \, ({\un x} -{\un y})}{dy} \Bigg|_{LO} \, = \, 
\frac{1}{4 \pi} \, \frac{z}{2} \, \sum_{\sigma, \lambda, a} 
\left< \psi^a_{\sigma\lambda} ({\un x}, z) 
\psi^{a*}_{\sigma\lambda} ({\un y}, z)  \right> \, = \, \frac{z \, N_c}{2} \, 
A \, \tas \, \ln \frac{1}{({\un x} -{\un y})^2 \Lambda^2}
\ee
where the brackets $\left< \, \ldots \, \right>$ imply averaging over
nucleon's impact factors and summation over all nucleons in the
nucleus \cite{kjklw,KM}, which brought in a factor of atomic number
$A$. We inserted a factor of $N_c$ in \eq{vqd1} to account for $N_c$
valence quarks in a nucleon and $\Lambda$ is some infrared
cutoff of the order of $\Lambda_{QCD}$. Since we are interested 
in the number of quarks per unit
rapidity we rewrote the $z$-integral in \eq{split} for small $z$ as
\ben
\frac{d z}{1-z} \, \approx \, d z \, = \, \frac{d z}{z} \, z
\een
which yielded the factor of $z$ in \eq{vqd1}. \eq{vqd1} gives us the
soft valence quark distribution in a nucleus at the lowest order in
the coupling constant. Its Fourier transform would give us
unintegrated valence quark distribution function of a nucleus.

Recalling that for soft gluons the corresponding lowest order wave
function is \cite{dip}
\be
A^a_{\sigma\lambda} ({\un x}_{21}, z) \, = \, 2 \, g \, T^a \,
\frac{i}{2 \pi} \, \frac{{\un \epsilon}^\lambda \cdot {\un
x}_{21}}{{\un x}_{21}^2}
\ee
we obtain the LO gluon distribution function \cite{kjklw,KM}
\be\label{glo}
\frac{dn_{G} \, ({\un x} -{\un y})}{dy} \Bigg|_{LO} \, = \, \frac{1}{4 \pi} 
\, \left< {\un A}^a ({\un x}) \cdot {\un A}^a ({\un y}) 
\right> \bigg|_{LO} \, = \, \tas \, A \, \ln \frac{1}{({\un x} -{\un y})^2 
\Lambda^2}.
\ee
Comparing \eq{vqd1} to \eq{glo} we see that already at this lowest
order the ratio of valence quarks to gluons in the small-$z$ tail of
the distribution functions is a rapidly falling function of rapidity
$y=\ln 1/z$
\be\label{r1}
\frac{dn_{val} /dy}{dn_G/dy} \Bigg|_{LO}  \, = \, 
\frac{z \, N_c}{2} \, \sim z \, \sim \, e^{-y}.
\ee
The exact scaling with rapidity of the ratio in \eq{r1} will be
modified by quantum evolution as we will see below.

In the mean time let us try to construct the valence quark
distribution function of a nucleus including the effects of multiple
rescatterings \cite{mv,kjklw,KM}. Diagrammatically this is equivalent
to resumming powers of $\as^2 A^{1/3}$, or, equivalently, powers of
$x_\perp^2 Q_s^2$ \cite{kjklw}. Following the prescription carried out
for gluonic fields in \cite{kjklw} we start off in $\partial_\mu A_\mu
=0$ covariant gauge. In this gauge nucleons can not exchange gluons
with each other and the soft gluonic or quark fields of the nucleus
are just additive with the total field of the nucleus being the sum of
the individual fields of the nucleons. Therefore if $\psi^{cov}_N
({\un x}, x_-)$ is the lowest order fermionic field of the valence
quarks in a single nucleon in covariant gauge the field of the whole
nucleus would be
\be\label{pcov}
\psi^{cov}_A ({\un x}, x_-) \, = \, \sum_{i=1}^A \psi^{cov}_{N_i} 
({\un x}, x_-).
\ee
After a gauge transformation to the $A_+ = 0$ light cone gauge the
field becomes
\be\label{lcpsi}
\psi^{LC}_A ({\un x}, x_-) \, = \, S ({\un x}, x_-) \, 
\sum_{i=1}^A \psi^{cov}_{N_i} 
({\un x}, x_-),
\ee
where the matrix of gauge transformation is \cite{kjklw,KM}
\be\label{S}
S ({\un x},x_-) \ = \ \mbox{P} \exp \left( - i g T^a \int \ d^2 b
\ d b_- \theta (x_- - b_-) \ {\hat \rho}^a ({\un b},b_-) \ \ln
(|{\un x} - {\un b}| \Lambda) \right).
\ee
${\hat \rho}^a$ is a color charge density operator normalized
according to
\be\label{dcorr}
\left< {\hat \rho}^a ({\underline x},x_-) \ {\hat \rho}^b ({\underline y},y_-) 
\right> \ = \ \frac{\as}{2 N_c \pi} \, \rho ({\underline x},x_-) \, 
\delta (x_- - y_-) \,  \delta^2 ({\underline x} - {\underline y}) \, 
\delta^{ab}
\ee
with $\rho ({\underline x},x_-)$ the normal nuclear density in the
infinite momentum frame of the nucleus, obeying
\be\label{dens}
\int \ d^2 x \ d x_- \ \rho ({\underline x},x_-) = \ A.
\ee
In terms of the fermionic field $\psi^{LC}_A ({\un x}, x_-)$ the
valence quark distribution function can be written as 
\be\label{vqd2}
\frac{d n_{val} \, ({\un x} -{\un y})}{dy} \, = \, \frac{1}{4 \pi} 
\, \frac{z}{2} \, \int \, 
\frac{d x_- \, d y_-}{(2 \pi)^2} \, e^{i k_+ (x_- - \, y_-)} \, \left<
{\bar \psi}^{LC}_A ({\un y}, y_-) \, \frac{1}{2} \gamma_+ \,
\psi^{LC}_A ({\un x}, x_-) \right> \nonumber\\
= \, \frac{1}{4 \pi} 
\, \frac{z}{2} \, \int \, 
\frac{d x_- \, d y_-}{(2 \pi)^2} \, e^{i k_+ (x_- - \, y_-)} \,
\left< {\bar \psi}^{cov}_{A} ({\un y}, y_-) \, S^{-1} ({\un y}, y_-)
\, \frac{1}{2} \gamma_+ \, 
S ({\un x}, x_-) \psi^{cov}_{A} ({\un x}, x_-) \right>,
\ee
where now $\left< \ldots \right>$ includes averaging over longitudinal
coordinates of the nucleons as well. The lowest order single nucleon
valence quark field $\psi^{cov}_N ({\un x}, x_-)$ is the same in light
cone and covariant gauges and is proportional to a $\delta$-function
on the light cone, $\psi^{cov}_{N_i} ({\un x}, x_-) \sim \delta (x_- -
x_{i-})$ with $x_{i-}$ the light cone coordinate of the nucleon. Using
this property of the fermion field together with \eq{pcov} in
\eq{vqd2} yields
\be\label{vqd2'}
\frac{d n_{val} \, ({\un x} -{\un y})}{dy} \, = \, \frac{1}{4 \pi} 
\, \frac{z}{2} \, \sum_{i=1}^A \, \left<
{\bar \psi}^{cov}_{N_i} ({\un y}, x_{i-}) \, S^{-1} ({\un y}, x_{i-})
\, \frac{1}{2} \gamma_+ \, S ({\un x}, x_{i-})
\psi^{cov}_{N_i} ({\un x}, x_{i-}) \right>.
\ee
To perform the averaging in \eq{vqd2'} one has to use the definition
of $S ({\un x}, x_-)$ from \eq{S} together with Eqs. (\ref{dcorr}) and
(\ref{dens}) in the product $S^{-1} ({\un y}, x_{i-}) S ({\un x},
x_{i-})$. Since the direction of $b_-$-integration is reversed in $S$
and $S^{-1}$ we can divide all the integration into tiny slices and
the first slice on the left of $S^{-1}({\un y}, x_{i-})$ would include
the same $b_-$ interval as the last slice on the right of $S ({\un x},
x_{i-})$. Since the color charge density correlators (\ref{dcorr}) are
local we can independently average in each slice keeping the
correlators only up to quadratic order in $\hat \rho$, which
corresponds to the quasi-classical approximation \cite{kjklw}. The
same procedure has been used for different correlators in
\cite{KM,kjklw}. Averaging of the $S$-matrices in \eq{vqd2'} can be done 
independent of ${\bar \psi} \gamma_+ \psi$ since the latter term is
the only one depending on the slice of the $b_-$-integral adjacent to
$x_{i-}$ (see the first reference in \cite{kjklw} for a discussion of
the ``last nucleon''). Performing $S$-matrix averaging \cite{KM} we
obtain
\be\label{vqd2''}
\frac{d n_{val} \, ({\un x} -{\un y})}{dy} \, = \, \frac{1}{4 \pi} 
\, \frac{z}{2} \, \sum_{i=1}^A \, \int_{-L}^L \, \frac{d x_{i-}}{2 \, L} 
e^{ - \frac{1}{4} ({\un x} -{\un y})^2 \, Q_s^{quark \, 2} \,  
\left( \frac{x_{i-} + L}{2 \, L} \right)} \left<
{\bar \psi}^{cov}_{N_i} ({\un y}, x_{i-}) \, \frac{1}{2} \gamma_+ \,
\psi^{cov}_{N_i} ({\un x}, x_{i-}) \right>,
\ee
where $2 \, L$ is the extent of the nucleus in the $x_-$-direction.
The quark saturation scale is defined for a cylindrical nucleus as
\cite{KT}
\be\label{xqs}
{\un x}^2 Q_{s}^{quark \, 2} \, = \, {\underline x}^2 \ \frac{2
\pi^2 \as}{N_c} \, \frac{A}{S_\perp} \, 
xG_N (x, 1/{\un x}^2)
\ee
with $S_\perp = \pi R^2$ the cross sectional area of the nucleus with
radius $R$ and $xG_N (x, 1/{\un x}^2) = \tas \ln (1/{\un x}^2
\Lambda^2)$ the gluon distribution in a single nucleon (see \eq{glo}).

Noting that the correlator in \eq{vqd2''} is, in fact,
$x_{i-}$-independent and is equal to the correlator in \eq{vqd1}, we
can integrate over $x_{i-}$ in \eq{vqd2''}. In the end we obtain the
following expression for the valence quark structure function of a
nucleus in a quasi-classical approximation
\be\label{vqd3}
\frac{d n_{val} \, ({\un x} -{\un y})}{dy} \, =  \, \frac{z \, N_c^2 \, 
S_\perp}{\as \, \pi^2 \, ({\un x} -{\un y})^2} \, 
\left( 1 - e^{- ({\un x} -{\un y})^2 Q_s^{quark \, 2} /4} \right).
\ee

\eq{vqd3} has an important qualitative feature in it: the valence
quark distribution goes to zero at large transverse separations $|{\un
x} -{\un y}|$. The effect of saturation physics on the valence quark
distribution function is to suppress the infrared part of the
distribution. This can be observed by defining the unintegrated
valence quark distribution function
\be\label{uvqd}
f_{val} (x=e^{-y} , {\un k}^2) \, = \, \frac{1}{4 \, \pi} \, \int d^2 x \,
e^{i {\un k} \cdot {\un x}} \,
\frac{d n_{val} \, ({\un x})}{d y}. 
\ee
Neglecting the ${\un x}$-dependence in $xG_N (x, 1/{\un x}^2)$ in
\eq{xqs} one can Fourier-transform the expression (\ref{vqd3}) getting
\be\label{kuvqd}
f_{val} (x , {\un k}^2) \, = \, \frac{z \, N_c^2 \, S_\perp}{4 \, \as
\, \pi^2} \ \Gamma \left(0, \frac{{\un k}^2}{Q_s^{quark \, 2}}
\right),
\ee
where $\Gamma(\nu,z)$ is the incomplete Gamma function. 
The valence quark distribution (\ref{kuvqd}) is much less infrared
divergent than the Fourier transform of the lowest order expression in
\eq{vqd1} (see \eq{uvqdlo} below). This property is similar to non-Abelian
Weizs\"{a}cker-Williams gluon distribution function \cite{kjklw}. One
can easily see that in the limit of small momenta $k_\perp \ll
Q_s^{quark}$ \eq{kuvqd} becomes
\be\label{irvqd}
f_{val} (x , {\un k}^2) \, = \, \frac{z \, N_c^2 \, S_\perp}{4 \, \as
\, \pi^2} \, \ln \frac{Q_s^{quark \, 2}}{{\un k}^2} \hspace*{1cm} 
k_\perp \ll Q_s^{quark}.
\ee
 At very high transverse momenta $k_\perp \gg Q_s^{quark}$ the valence
quark distribution function (\ref{vqd3}) transformed via \eq{uvqd}
maps onto $f_{val} (x , {\un k}^2)$ given by the lowest order
expression which can be obtained by either using \eq{vqd1} in
\eq{uvqd} or by expanding \eq{kuvqd}
\be\label{uvqdlo}
f_{val} (x , {\un k}^2) \bigg|_{LO} \, = \, \frac{z \, N_c}{2} \, A \,
\tas \, \frac{1}{{\un k}^2}, \hspace*{1cm} k_\perp \gg Q_s^{quark}.
\ee

The quasi-classical approximation employed in this Section allowed us
to derive two important features of the valence quark distribution at
small-$x$. The first feature is that, similar to the gluon
distribution, multiple rescatterings regulate the infrared singularity
as demonstrated in Eqs. (\ref{vqd3}), (\ref{kuvqd}) and
(\ref{irvqd}). The second feature is that rapidity/Bjorken $x$
distribution of valence quarks at the classical level is just
proportional to $x$ (see \eq{r1}). In the following three Sections we
will study how this $x$-dependence gets modified by inclusion of
quantum evolution.

\section{Including nonlinear evolution in the 
double logarithmic approximation}
\label{QUANTUM}

Our goal here is to describe the small-$x_{Bj}$ evolution of the
valence quark distribution functions including the effects of gluon
evolution to all orders in color charge density. We consider deep
inelastic scattering (DIS) on a nucleus.  In the dipole picture
\cite{dip,dip1} the splitting function of a virtual photon into
$q\bar{q}$ dipole is factorized from the subsequent evolution of the
scattering cross section of the dipole on a nucleus \cite{yuri}.  The
total cross section and $F_2$ structure function are dominated by
gluon exchange at small $x_{Bj}$ for which one writes
\be\label{f2tot}
 F_2(x_{Bj}, Q^2) = \frac{Q^2}{4 \pi^2 \alpha_{EM}} \int \frac{d^2 x \, d
 z }{4 \pi} \, \Phi^{\gamma^* \rightarrow q{\bar q}} ({\un x}, z) \
 2 \int d^2 b \ 
 N({\un x},{\un b}, \tau \equiv \ln \frac{z_{\min}s}{\Lambda^2} ).
\ee
Here
 $\Phi({\un x},z)$ describes the photon splitting into a
$\bar{q}q$ dipole with transverse separation $\un{x}$  and  
moment fractions $z$ and $1-z$ respectively.
To first order in the electromagnetic charge (see e.g. \cite{YL}) 
the photon splitting 
function into a $\bar{q}q$ dipole, with flavor $f$ and 
electromagnetic charge $(Z_fe)$  is
\begin{equation}
\label{photonsplit}
   \Phi^{\gamma^{*}\rightarrow q\bar{q}}_f
   (\un{x},z) = 
   \frac{2\alpha_{EM}\ Z_f^2  N_c} {\pi} 
   \left[ \epsilon^2 K_1^2(\epsilon x) \left(z^2 + (1-z)^2\right)
   + 4\, Q^2\, z^2(1-z)^2\,K_0^2(\epsilon x) \right] \;.
\end{equation}
where $\epsilon^2=z(1-z)Q^2 +m_f^2$ and $m_f$ is the quark mass.
\eq{f2tot} implicitly includes a sum over all quark flavors.  $N({\un
x},{\un b},\tau)$ is the forward scattering amplitude at impact
parameter $\un{b}$ of the $q\bar{q}$ dipole on a
nucleus. $N(\un{x},{\un b},\tau)$ depends upon the rapidity $\tau =
\ln \frac{z_{\min} s}{\Lambda^2}$ where $s = \frac{Q^2}{x_{Bj}}$ and
$z_{\min}=\min(z,1-z)$. In the leading log approximation we substitute
$\tau = \ln \frac{zs}{\Lambda^2} \approx \ln\frac{1}{x_{Bj}}$ provided
that $z$ is not too small.  $N({\un x},{\un b},\tau)$ is normalized
such that
\be\label{norm}
\sigma^{q{\bar q} A}_{glue} \, = \, 2 \int d^2 b \, 
 N({\un x},{\un b} , \tau).
\ee

The quantum evolution of the gluon exchange amplitude $N({\un x},{\un
b}, \tau)$ has been resummed to all orders in pomeron exchanges/color
charge density \cite{yuri,dip,bal,JKLW,FILM} for the total cross
section of a $q\bar q$ dipole scattering on a nucleus. The result was
a functional differential equation \cite{JKLW,FILM}. In the large
$N_c$ limit the small $x_{Bj}$ evolution equation for the forward
amplitude $N({\un x}, {\un b}, \tau)$ is a nonlinear
integro-differential equation \cite{yuri,bal}
\ben
  N({\underline x}_{01},{\underline b}, \tau) = \gamma ({\underline
  x}_{01},{\underline b}) \, \exp \left[ - \frac{4 \as C_F}{\pi} \ln
  \left( \frac{x_{01}}{\rho} \right) \tau \right] + \frac{\as
  C_F}{\pi^2} \int_0^\tau d \tau' \, \exp \left[ - \frac{4 \as C_F}{\pi}
  \ln \left( \frac{x_{01}}{\rho} \right) (\tau - \tau') \right]
\een
\be\label{eqN}
\times \int_\rho d^2 x_2 \frac{x_{01}^2}{x_{02}^2 x_{12}^2} \, [ 2
  \, N({\underline x}_{02},{\underline b} + \frac{1}{2} {\underline
  x}_{21}, \tau') - N({\underline x}_{02},{\underline b} + \frac{1}{2}
  {\underline x}_{21}, \tau') \, N({\underline x}_{12},{\underline b} +
  \frac{1}{2} {\underline x}_{20}, \tau') ] ,
\ee
where $\rho$ is some ultraviolet cutoff  and
$\gamma ({\underline x}_{01},{\underline b})$ is the initial
condition for $\tau$-evolution. In Ref.~\cite{yuri} 
$\gamma ({\underline x}_{01},{\underline b})$ 
was taken in the
quasi-classical Mueller-Glauber approximation \cite{mugla}
\begin{equation}\label{gla}
   \gamma ({\underline x},{\underline b}_0) =1- e^{ - {\underline x}^2
   Q_{s}^{quark 2} / 4} \;.
\end{equation}
The linear part of \eq{eqN} gives the BFKL equation
\cite{BFKL,dip}. Inclusion of multiple pomeron exchanges introduces
the term quadratic in $N$ on the right hand side of \eq{eqN}.  In
momentum space at $t=0$ and in the double logarithmic limit \eq{eqN}
reduces to GLR-MQ equation
\cite{glrmq}. We want to construct an analogue of \eq{eqN} for the
valence quark distribution function.

To this end, consider the difference between the structure functions
of a nucleus made entirely of protons and a nucleus made entirely of
neutrons, $F_2^{val} \equiv 3 (F_2^{p} - F_2^{n})$.  The gluon
exchange amplitudes are identical for the proton and neutron and
therefore $F_2^{val}$ depends upon amplitudes in which a valence quark
is exchanged between the photon and the target.  As before, we first
factor the photon splitting function from the subsequent interaction
of the $q\bar{q}$ dipole with the nucleus. Then we define the forward
scattering amplitude of a dipole on a nucleus interacting with a
single $\bar{q}$ exchange, $R({\un x}, {\un b}, z_1)$, where $z_1$ is
the light-cone momentum fraction carried by the anti-quark.  As in
\eq{f2tot}, the valence quark structure function can be obtained from
$R({\un x}, {\un b}, z_1)$ by convoluting $R$ with the photon
splitting function
\be\label{f2}
  F_2^{val} (x_{Bj}, Q^2) = \frac{Q^2}{4 \pi^2 \alpha_{EM}} \int \frac{d^2
  x \, d z }{4 \pi} \, \Phi^{\gamma^* \rightarrow q{\bar q}} ({\un x}, z)
  \ 2 \int  d^2 b \ R({\un x},{\un b}, z_1) \;.
\ee
It is helpful to imagine scattering a dipole made of strange quarks
(i.e. $s\bar{s}$)  on a ``nucleus'' made up of $u\bar{d}$  dipoles in
a world with three massless flavors.
Then the difference between the $u\bar{u}$  and
$s\bar{s}$ dipole cross sections is simply given by the reggeon exchange 
amplitude
\begin{equation}
   \sigma_{val}^{q\bar{q}} \equiv \sigma_{tot}^{u\bar{u}A} - 
   \sigma^{s\bar{s}A}_{tot}
   = 2 \int d^2b \ R(\un{x}, \un{b}, z_1) \; .
\end{equation}
In the quasi-classical and double logarithmic approximations
considered below only the antiquark in the original dipole will be
responsible for the flavor exchange interaction with the nucleus.  The
amplitude $R$ will therefore depend on the antiquark momentum fraction
$z_1$ only.

\begin{figure}
\begin{center}
\epsfxsize=8cm
\leavevmode
\hbox{ \epsffile{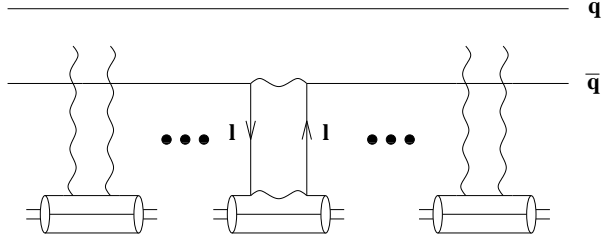}}
\end{center}
\caption{Forward amplitude of a $q\bar q$ pair interaction with the 
nucleus with one flavor-exchange interaction and all orders in gluon
exchange rescatterings.}
\label{init}
\end{figure}

To derive an evolution equation for $R(\un{x},\un{b},z)$
we first need to construct the quasi-classical
initial conditions. The diagrams we need to sum for that are shown in
\fig{init}. For gluonic evolution (\ref{eqN}) the initial condition 
(\ref{gla}) is obtained by summing up multiple rescatterings of the
dipole on the nucleons in the nucleus with two gluons exchanged with
each interacting nucleon \cite{mugla}. To construct initial conditions
for the valence quark distribution function we modify that by
replacing {\sl one} of the gluonic exchanges by the quark exchange
amplitude (see \fig{init}). Each quark exchange would bring in a
suppression factor of $1/s$ with $s$ the center of mass energy. We
therefore restrict ourselves to the leading term in this $1/s$
expansion.  For simplicity, we shall make all the quarks in the
nucleus and the dipole a single flavor and therefore there are $N_c A$
valence quarks in the nuclear wave function which can be exchanged
with the dipole.

To calculate the diagram shown in \fig{init}
we start by considering only the $ q{\bar q}$ exchange part.
A simple calculation given in Appendix \ref{qbarq} yields
\be\label{qq}
 \sigma^{q\bar{q}}_{val} = \frac{2 \, \as^2 \, C_F^2}{N_c \, z_1 \, s} \, \int \, \frac{d^2 l}{{\un l}^2}
\ee
where $s = (p+q)^2$ is the center of mass energy per nucleon with the
virtual photon carrying momentum $q$ and the nucleon having momentum
$p$. The integration over transverse momentum $l$ of the exchanged
quarks in \eq{qq} is logarithmically divergent both in the infrared
and in the ultraviolet. We will cut it off from below by the same
infrared cutoff $\Lambda^2$ that we have used in the previous Section
and we will cut it off from above by the maximum available relevant
momentum scale in the problem --- the center of mass energy $z_1 s$ of
the antiquark--nucleon system. Thus \eq{qq} becomes
\be\label{qqq}
 \sigma^{q\bar{q}}_{val} = \frac{2 \, \as^2 \, C_F^2 \, \pi}{N_c \, z_1 \, s} \, 
\ln \frac{z_1 \, s}{\Lambda^2}.
\ee
The effect of multiple two gluon exchanges with different nucleons 
is easy to calculate and gives a factor of (see e.g. \cite{YL}) 
\be
e^{- {\un x}^2 Q_s^{quark \, 2} /4}\; . 
\ee
This factor is understood easily in the context of Glauber theory as
$e^{-\frac{1}{2}n \sigma L}$, where $n$ is the nucleon density,
$\sigma$ is dipole nucleon cross section, and $L$ is the path length
of the dipole in the nucleus. The factor of $\frac{1}{2}$ arises
because we are calculating the amplitude while it is the amplitude
squared which gives the dipole survival probability $e^{-n\sigma L}$.
The forward amplitude $R$ in the quasi-classical approximation is then
\be\label{incond}
R_0 ({\un x},{\un b}, z_1) \, = \, \frac{ \as^2 \, C_F^2 
\, \pi}{ z_1 \, s} \, \frac{A}{S_\perp} \, \ln 
\frac{z_1 \, s}{\Lambda^2} \, 
e^{- {\un x}^2 Q_s^{quark \, 2} /4},
\ee
Here a factor $(N_c A)/S_\perp$ has been inserted since the 
antiquark can be exchange with any of the  $(N_c A)/S_\perp$ quarks 
at a given impact parameter.
\begin{figure}
\begin{center}
\epsfxsize=6cm
\leavevmode
\hbox{ \epsffile{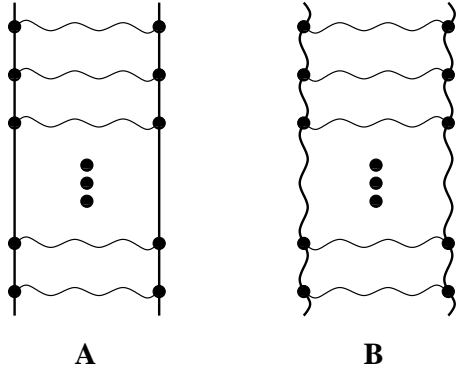}}
\end{center}
\caption{(A) Ladder diagram with the reggeized quarks in the 
$t$-channel and effective quark-gluon vertices considered in 
\protect\cite{Kirschner1}; (B) Standard BFKL ladder diagram with 
reggeized gluons in the $t$-channel and effective Lipatov vertices.}
\label{poms}
\end{figure}

Now that we have constructed the initial condition we may start
building up the small-$x$ evolution. Linear equation for a BFKL-like
ladder with quarks instead of gluons in the $t$-channel (see
\fig{poms}A) has been originally constructed in \cite{Kirschner1} and
was subsequently studied in \cite{Kirschner2,GR,EMR}. The resulting
evolution equation has a peculiar property: due to the presence of
quarks in the $t$-channel a certain part of the transverse momentum
integration in the kernel is logarithmically divergent, similar to the
integral in \eq{qq}. The integration is effectively cut off in the
ultraviolet by the center of mass energy giving an extra $\ln s$ per
rung of the ladder \cite{Kirschner1}. Therefore the ladder of
\fig{poms}A with quarks in the $t$-channel at the leading order in
$\ln s$ effectively resums powers of the parameter $\as \ln^2 s$. This
resummation is usually referred to as double logarithmic approximation
(DLA). Note that this is in contrast to the usual BFKL ladder
\cite{BFKL} (see \fig{poms}B) which resums powers of $\as \ln s$ and 
for which DLA implies resummation of $\as \ln s \ln Q^2$. Of course
the quark ladder also has an UV-safe part of the kernel resummation of
which gives powers of $\as \ln s$ \cite{Kirschner2,GR}. However, it
seems a little unclear whether resummation of higher order corrections
to the DLA part of the kernel would not give contributions
parametrically of the same size as iterations of the UV-safe part of
the kernel. For instance NLO correction to the DLA kernel would give a
parametric factor of $\as^2 \ln^2 s$, which is equivalent to double
iteration of the leading logarithmic UV-safe part of the kernel $(\as
\ln s)^2$. In what follows below we will work only in the double
logarithmic approximation for the reggeon amplitude evolution to avoid
these complications which may potentially require resummation of
perturbation theory to all orders to find the leading logarithmic
contribution.

At the same time we want to resum all multiple BFKL pomeron exchanges
in gluon evolution similar to how it was done in
\cite{yuri}. Each pomeron is taken in the leading logarithmic 
approximation giving a parametric contribution of the order $\as \, \rho
\, e^{C \as \ln s}$, where $\rho$ is proportional to the color charge 
density of the nucleus \cite{mv,kjklw} and is large ($\rho \sim \as \,
A^{1/3}$). Therefore, even though gluon evolution will be taken below
in the leading logarithmic approximation while reggeon evolution will
be in the DLA limit, one can see that pomeron exchanges of the gluon
evolution are enhanced by the powers of the color charge density of
the source while the leading logarithmic part of reggeon evolution is
not. It is therefore justified to include gluon evolution in the
leading logarithmic approximation enhanced by color charge density
while keeping only the DLA part of the reggeon evolution.

We want to construct an analogue of Mueller's dipole model
\cite{dip} for the reggeon amplitude $R$. To construct a dipole 
evolution one has to take 't Hooft's large $N_c$ limit \cite{th} in
the dipole's wave function. The advantage of this approach is that
inclusion of gluon evolution effects would become straightforward
\cite{yuri}.

\begin{figure}
\begin{center}
\epsfxsize=15cm
\leavevmode
\hbox{ \epsffile{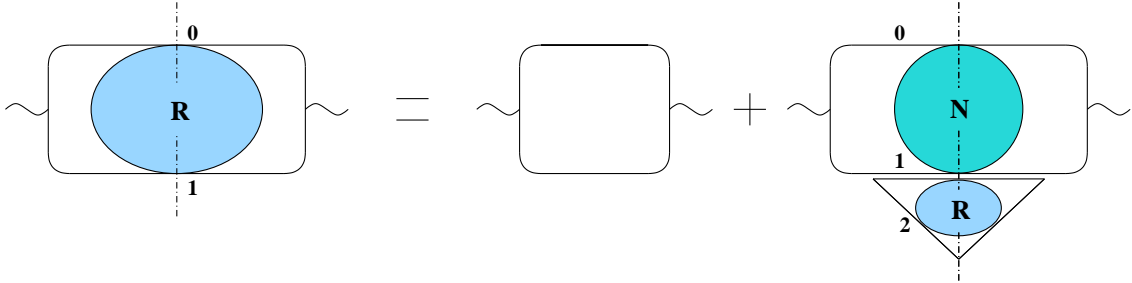}}
\end{center}
\caption{Evolution equation for the reggeon amplitude 
$R({\un x}_{01}, {\un b}, z_1)$ in the double logarithmic
approximation.  As defined above, $R({\un x}_{01}, {\un b}, z_1)$
is the forward amplitude of a $q\bar q$ dipole interacting with the
nucleus by a single $\bar q$ exchange and many gluon exchanges, while
$N({\un x}_{01}, {\un b}, z_1)$ is the forward amplitude of a $q\bar
q$ dipole interacting with the nucleus by gluon exchanges only.}
\label{dla}
\end{figure}

A single step of the evolution is the same as shown in \fig{lo}. A
hard valence quark splits into a soft quark and a hard gluon. A new
color dipole $23$ is created \cite{dip}. To iterate this kernel one
can write down an evolution equation which is depicted in
\fig{dla}. There an initial dipole $01$ either has no evolution 
in it at all (the first term on the right hand side in \fig{dla}) or
the antiquark $1$ in the dipole splits into a gluon (double line) and
an antiquark $2$ creating a new dipole $12$ in addition to preexisting
dipole $01$. Since we are iterating the kernel in which the antiquark
$2$ is always much softer than the antiquark $1$ the transverse
coordinate of the antiquark $2$ is the same on both sides of the
cut. The subsequent reggeon evolution can continue in the dipole $12$
while usual gluon evolution may happen in the dipole $01$.  Since in
the double logarithmic approximation the virtual corrections are not
important (since they do not give two logarithms of energy $s$) we can
write down an equation using only the real part of the kernel (see
e.g. \cite{glrmq,yuri}).
Using the kernel of \eq{split} with $z=z_2/z_1 \ll 1$ we write
\ben
R({\un x}_{01}, {\un b}, z_1) \, = \, R_0({\un x}_{01}, {\un b},
z_1) \, + \, \frac{\tas}{2 \, \pi} \, \int_{z_i}^{z_1 \, \mbox{min}
\{1, x_{01}^2/x_{21}^2 \}} \, \frac{d z_2}{z_1} \, \frac{d^2
x_2}{x_{21}^2}
\, R({\un x}_{12}, {\un b} + \frac{1}{2} {\un x}_{20}, z_2) \, 
\een
\be\label{eqr1}
\times \, \left[1 - N({\un x}_{01}, {\un b}, z_1)\right]
\ee
where we have switched from rapidity notation to momentum fraction
$z_1$ in the argument of $N$ as well. ($N$ depends on the rapidity of
the softer quark or antiquark in the dipole \cite{yuri,dip}, which in
the case of \eq{eqr1} is the antiquark rapidity determined by $z_1$. ) 
The first term on the right hand side of \eq{eqr1} corresponds to the
first term on the right hand side in \fig{dla} and represents the
initial conditions given by \eq{incond}. The transverse coordinate
integral in the second term on the right hand side of \eq{eqr1} is
logarithmically divergent so one has to take extra care to define the
limits of the $z_2$-integration to make it finite
\cite{Kirschner1,Kirschner2,GR}. This property of the valence quark
distribution's evolution makes it very different from the usual
gluonic evolution. Our goal is to order dipoles in rapidity so that
each newly formed dipole would have smaller rapidity with respect to
the target than the previous dipole it was produced in. In the double
logarithmic approximation we can assume that the antiquark in a dipole
always carries a softer light cone momentum than the quark. Defining
the rapidity of dipole $01$ as $Y = \ln (z_1 s
\, x_{01}^2)$ we would get $y = \ln (z_2 s \, x_{12}^2)$ for the 
rapidity of dipole $12$. (Dipole $01$ after splitting has roughly the
same rapidity as before the splitting. Details of rapidity ordering
discussed here for quarks are not important for the gluon evolution
$N$.) Requiring that $Y \gg y$ we get 
\be\label{req1}
z_2 \ll z_1 \, \frac{x_{01}^2}{x_{21}^2},
\ee
which can also be obtained by requiring that the soft valence quark
contribution dominates in the energy denominator. On the other hand in
order to obtain logarithm of energy we need the ${\bar q} \rightarrow
G {\bar q}$ splitting to produce a soft quark (see \eq{split}). This
translates into requirement that
\be\label{req2}
z_2 \ll z_1.
\ee
Combining Eqs. (\ref{req1}) and (\ref{req2}) we end up with
\be
z_2 \ll z_1 \, \mbox{min} \{ 1, \, \frac{x_{01}^2}{x_{21}^2}\}
\ee
which is the upper limit used in $z_2$ integral in \eq{eqr1}.

\eq{eqr1} together with \eq{f2} provide us with the way of determining 
the valence quark distribution function if the forward dipole
amplitude $N$ has been found previously from \eq{eqN}. Defining
\be\label{redef}
\tilde{R} ({\un x}_{01}, {\un b}, z_1) \, \equiv \, z_1s \, R 
({\un x}_{01}, {\un b}, z_1)
\ee
we can rewrite \eq{eqr1} as
\ben
\tilde{R}({\un x}_{01}, {\un b}, z_1) \, = \, \tilde{R}_0
({\un x}_{01}, {\un b},
 z_1) \, + \, \frac{\tas}{2 \, \pi} \, \int_{z_i}^{z_1 \,
\mbox{min} \{1, x_{01}^2/x_{21}^2 \}} \, \frac{d z_2}{z_2} \, \frac{d^2
x_2}{x_{21}^2} \, \tilde{R} ({\un x}_{12}, {\un b} + \frac{1}{2} 
{\un x}_{20}, z_2) \, 
\een
\be\label{eqr2}
\times \, \left[1 - N({\un x}_{01}, {\un b}, z_1)\right]. 
\ee
\eq{eqr2} explicitly demonstrates that the $z_2$ integral of \eq{eqr2} is 
logarithmic and is similar in structure to the equation describing jet
decays in the modified leading logarithmic approximation (MLLA)
\cite{DKT}.

 One can determine the scale for the running coupling constant in
\eq{eqr2} following the prescription outlined in \cite{DS}. Since the 
lowest order $\as N_c$ and $\as N_f$ running coupling corrections
should come together to give leading order QCD beta-function, we need
to calculate only the $\as N_f$ terms to obtain the answer ($N_f$ if
the number of flavors). Writing a dispersion relation for the gluon
propagator in \fig{lo} one can show that the scale for the strong
coupling constant is given by $({\un k} - z {\un p})^2 /z$, such that
$\as = \as [({\un k} - z {\un p})^2 /z]$ \cite{DS}. In the small-$z$
limit used to obtain \eq{eqr2} this reduces to $\as ({\un k}^2
/z)$. In the coordinate space for one step of the evolution
(\ref{eqr2}) this translates into $\as [z_1 /(z_2 \, x_{21}^2)
]$. Therefore, running coupling effects can be included in \eq{eqr2}
by replacing $\tas$ in front of the integral by $\tas [z_1 /(z_2 \,
x_{21}^2) ]$ in the integrand. 
As we will see in Sect. V, the evolution equation (\ref{eqr1}) cuts off 
the infrared region with momenta less than $Q_s$ in the reggeon amplitude 
$R$. Similar absence of infrared diffusion was observed previously 
for the nonlinear evolution equation (\ref{eqN}) in \cite{GMS}. 
Therefore, together with the above argument, 
the scale of the coupling constant can, practically, 
only be set by either $Q_s$ or a higher momentum. In either case 
the coupling would be small for parametrically large energies. 
In the following, we use simply $Q_s$ for the running coupling scale.

\section{Solution of the linear DLA equation}
\label{linear}

Let us check that the linear part of \eq{eqr2} is consistent with the
DLA piece of the equation derived in \cite{Kirschner1} by explicitly
solving it and comparing the result with
\cite{Kirschner2,GR,EMR}. Outside of the saturation region $N \ll 1$
\cite{LT,IIM,yuri} and we can neglect it on the right hand side of 
\eq{eqr2} obtaining
\be\label{lineq}
\tilde{R}({\un x}_{01},  z_1) \, = \, \tilde{r}_0 ({\un x}_{01}, z_1) 
\, + \, \frac{\tas}{2 \, \pi} \, \int_{z_i}^{z_1 \,
\mbox{min} \{1, x_{01}^2/x_{21}^2 \}} \, \frac{d z_2}{z_2} \, \frac{d^2
x_2}{x_{21}^2} \, \tilde{R} ({\un x}_{12}, z_2)
\ee
where $\tilde{r}_0 ({\un x}_{01}, z_0, z_1)$ is obtained from
\eq{incond} by putting the exponent to be equal to $1$ so that
\be\label{r0}
\tilde{r}_0 ({\un x}_{01}, z_1) \, = \,  
\as^2 \, C_F^2 \, \pi \, \frac{A}{S_\perp} \, \ln 
\frac{z_1 \, s}{\Lambda^2} \, = \, \tilde{r}^{(0)} \, \left( Y  + 
\ln \frac{1}{x_{01}^2 \, \Lambda^2} \right)
\ee
with
\be\label{r00}
\tilde{r}^{(0)} \, = \,  
\as^2 \, C_F^2 \, \pi \, \frac{A}{S_\perp}.
\ee
In \eq{lineq} we suppressed the impact parameter dependence neglecting
the shift in ${\un b}$ on the right hand side which is a good
approximation for a central collision of a $q\bar q$ pair with a large
nucleus \cite{LT,IIM,yuri}.  Introducing Laplace transform in rapidity
\be\label{lap}
\tilde{R}({\un x}_{01}, z_1) \, = \, \int 
\frac{d \omega}{2 \pi i} \, e^{\omega Y} \, \tilde{R}_\omega ({\un x}_{01}) 
\, = \, \int \frac{d \omega}{2 \pi i} \, (z_1 s \, x_{01}^2)^\omega \, 
\tilde{R}_\omega ({\un x}_{01})
\ee
we rewrite \eq{lineq} as
\be\label{lineq2}
\omega \, \tilde{R}_\omega (x_{01}) \, = \, \left( \frac{1}{\omega}  + 
\ln \frac{1}{x_{01}^2 \, \Lambda^2} \right) \, 
\tilde{r}^{(0)} \, + \, 
\frac{\tas}{2} \, \int_0^\infty \frac{d
x_{21}^2}{x_{21}^2} \, \left( \mbox{min} \left\{
\frac{x_{21}^2}{x_{01}^2}, 1 \right\}\right)^\omega \, 
\tilde{R}_\omega (x_{21})
\ee
where we are interested only in the azimuthally symmetric solution
which is dominant at high energy. Note that we implicitly assume that
rapidity $Y > 0$, so that $x_{01}^2 > 1/(z_1 \, s)$ in the first term
on the right hand side of \eq{lineq}. Then $\omega$-integration in
\eq{lap} runs parallel to imaginary axis to the right of the origin.
Defining Mellin transform
\be\label{mel}
\tilde{R}_\omega ({\un x}_{01}) \, = \, \int \frac{d \lambda}{2 \pi i} \, 
(x_{01}^2 \Lambda^2)^{-\lambda} \, \tilde{R}_{\omega \lambda}
\ee
reduces \eq{lineq2} to
\be
\omega \, \tilde{R}_{\omega \lambda} \, = \, 
\frac{\omega + \lambda}{\omega \lambda^2} \, 
\tilde{r}^{(0)} \, + \, \frac{\tas}{2} \, \frac{\omega}{\lambda 
(\omega - \lambda)} \, \tilde{R}_{\omega \lambda}
\ee
which gives
\be\label{sol1}
 \tilde{R}_{\omega \lambda} \, = \, \frac{\tilde{r}^{(0)} \, (\omega^2
 - \lambda^2)}{\omega^2 \, \lambda \, (\omega \lambda - \lambda^2 -
 \frac{\tas}{2})}.
\ee
Combining Eqs. (\ref{sol1}), (\ref{mel}) and (\ref{lap}) yields
\be\label{sol2}
\tilde{R}({\un x}_{01}, z_1) \, = \, \int \frac{d \omega}{2 \pi i} 
\, \frac{d \lambda}{2 \pi i} \, (z_1 s \, x_{01}^2)^\omega \,  
(x_{01}^2 \Lambda^2)^{-\lambda} \, \frac{\tilde{r}^{(0)} \, (\omega^2
- \lambda^2)}{\omega^2 \, \lambda \, (\omega \lambda - \lambda^2 -
\frac{\tas}{2})}.
\ee
Performing the $\omega$-integration first in \eq{sol2} we notice that
for positive $Re \, \lambda$ the high energy asymptotics is dominated
by the rightmost pole
\be
\omega \, = \, \omega^* (\lambda) \, \equiv \, \lambda + \frac{\tas}{2 \, 
\lambda}.
\ee
\eq{sol2} becomes
\be\label{sol3}
\tilde{R}({\un x}_{01}, z_1) \, = \, \int  
\frac{d \lambda}{2 \pi i} \, (z_1 s \, x_{01}^2)^{\omega^* (\lambda)} \,  
(x_{01}^2 \Lambda^2)^{-\lambda} \, \frac{\tilde{r}^{(0)} \, (\omega^{* 2}
(\lambda) - \lambda^2)}{\lambda^2 \, \omega^{* 2} (\lambda)} .
\ee
The integral in \eq{sol3} can be done in the saddle point
approximation around the saddle point $\lambda^* = \sqrt{\tas /2}$
yielding
\be\label{sol4}
\tilde{R}({\un x}_{01}, z_1) \, = \, \frac{3 \, \tilde{r}^{(0)}}{4 \, 
\sqrt{\pi}} \, \left( \frac{2}{\tas} \right)^{1/4} \, 
(x_{01} \, \Lambda)^{- \sqrt{2 \, \tas}} \, \frac{e^{\sqrt{2 \, 
\tas} \, Y} }{\sqrt{2 \, \tas \, Y}} \ e^{-\frac{\sqrt{\tas}}{Y \sqrt{2}} 
\, \ln^2 (x_{01} \Lambda)}
\ee
where we switched to rapidity notation with $Y = \ln (z_1 s \,
x_{01}^2)$. The integral in \eq{sol3} can, in fact, be done almost
exactly as will be shown in the next Section for a more general
case. Using \eq{sol4} in \eq{redef} together with \eq{r00} gives
\be\label{sol5}
R({\un x}_{01}, z_1) \, = \, \frac{3}{4} \, \tas^2 \, \pi^2 \,
\sqrt{\pi} \, \frac{A}{S_\perp} \, 
 \left( \frac{2}{\tas} \right)^{1/4} \, 
x_{01}^2 \, (x_{01} \, \Lambda)^{- \sqrt{2 \, \tas}} \, \frac{e^{(\sqrt{2 \, 
\tas} - 1) \, Y} }{\sqrt{2 \, \tas \, Y}} \ e^{-\frac{\sqrt{\tas}}{Y \sqrt{2}} 
\, \ln^2 (x_{01} \Lambda)}
\ee
The intercept of the reggeon in \eq{sol5} is equal to
\be
\alpha_R \, = \, \sqrt{2 \, \tas}
\ee
in agreement with \cite{KL,Kirschner1,Kirschner2,GR,EMR} so that
\be
R \, \sim \, e^{(\alpha_R -1) Y}\, \sim \, e^{(\sqrt{2 \, \tas} - 1)
\, Y}.
\ee 
To understand the transverse coordinate dependence induced by
evolution let us rewrite the initial conditions of \eq{incond} in
terms of rapidity. A simple calculation yields
\be\label{incond1}
R_0 ({\un x},{\un b}, z_1) \, = \,  \as^2 \, C_F^2 
\, \pi \, \frac{A}{S_\perp} \, x_{01}^2 \, e^{-Y} \, \left(Y +  \ln 
\frac{1}{x_{01}^2 \Lambda^2} \right) \, 
e^{- {\un x}^2 Q_s^{quark \, 2} /4}.
\ee
Comparing \eq{incond1} to \eq{sol5} we see that the transverse
coordinate dependence of $R$ gets modified by the factor of $(x_{01}
\, \Lambda)^{- \sqrt{2 \, \tas}}$ which also agrees with the results of
\cite{KL,Kirschner1,Kirschner2,GR,EMR}. We see that the small-$x$
evolution makes the valence quark distribution even more sensitive to
the ultraviolet region by pushing the quarks toward higher transverse
momenta. Overall we conclude that we have constructed a solution of
the linear part of \eq{eqr1} shown in \eq{sol5} and found it to be in
agreement with the previous studies of the perturbative reggeon
\cite{KL,Kirschner1,Kirschner2,GR,EMR}.

\section{Solution of the Full Nonlinear Equation with a Simple Model 
for $N$}
\label{nonlinearSection}

Let us consider a simple model for the for the forward dipole
amplitude $N({\underline x}_{01},{\underline b}, \tau)$ which has correct
qualitative features in agreement with the solution of \eq{eqN}
\cite{yuri,bal} obtained in \cite{LT,IIM}. Let us take
\be\label{modelN}
N({\underline x}_{01},{\underline b}, \tau) \, = \, \theta ({\underline
x}_{01}^2 Q_s^2 (\tau) - 1) \; .
\ee
The saturation scale $Q_s$ changes with energy as
\be
Q_s^2 \, = \, \Lambda^2 \, e^{\kappa \tau} \;,
\ee
where
\be
\tau \, = \, \ln \frac{z \, s}{\Lambda^2} \; ,
\ee
and $\kappa$ is the intercept which is usually taken to be $0.2 - 0.3$
\cite{LT,IIM,MT}. \eq{modelN} gives us the dipole amplitude $N$ which is 
equal to one inside the saturation region and is zero otherwise. In the
spirit of the theta-function approximation we also model the initial
conditions of \eq{eqr2} given in \eq{incond} by
\be\label{modelR0}
{\tilde R}_0 ({\underline x}_{01}, \tau) \, = \, {\tilde r}^{(0)} \,
\tau \, \theta (1 - {\underline x}_{01}^2 Q_s^2 (\tau)). 
\ee
Plugging Eqs. (\ref{modelN}) and (\ref{modelR0}) into \eq{eqr2} we
immediately see that the solution for ${\tilde R} ({\underline
x}_{01}, z)$ is zero inside the saturation region and can therefore be
parameterized as
\be\label{R1}
{\tilde R} ({\underline x}_{01}, z) \, = \, \theta (1 - {\underline
x}_{01}^2 Q_s^2 (\tau)) \, {\tilde R}_1 ({\underline x}_{01}, z),
\ee
where we again suppress the impact parameter dependence.  Substituting
\eq{R1} together with Eqs. (\ref{modelN}) and (\ref{modelR0}) into
\eq{eqr2} we obtain
\be\label{oreq}
{\tilde R}_1 (\eta, Y) \, = \, {\tilde r}^{(0)} \, p \, (Y+\eta) +
\frac{\tas \, p}{2} \, \int_0^{Y} d Y' \,
\int_0^{\eta + Y - Y'} d \eta' \, {\tilde R}_1 (\eta', Y')
\ee
where we have defined
\be
\eta = \ln \frac{1}{x_\perp^2 Q_s^2} 
\ee
and made use of the fact that 
\be
Y \, = \, \ln (z \, s \, x_\perp^2 ) \, = \,  \frac{\tau}{p} - \eta
\ee
with
\be
p \equiv \frac{1}{1-\kappa}.
\ee
For simplicity we have also assumed that
\be
\tau_{in} \, \equiv \, \ln \frac{z_i \, s}{\Lambda^2} \, = \, 0.
\ee
Note that $\eta, Y >0$. 

To solve \eq{oreq} we reduce it to partial
differential equation, determine the boundary conditions
from the original integral equation, and then solve 
the differential equation using Laplace transform methods. This 
is done in Appendix~\ref{solution}. The complete solution
is 
\begin{equation}
\label{sol}
{\tilde R}_1 (\eta, Y)\,=\, 
\frac{ 
          \tilde{r}^{(0)} p 
        } {\gamma} \,
    \left[ \frac{Y+\eta} 
                { \sqrt{Y(Y+\eta)}  } \,
                I_1(2 \gamma  \sqrt{Y(Y+\eta)})
               - 
           \frac{ (Y(Y+\eta))^{3/2} }
                       {(Y+\eta)^3}  \, 
                I_{3}(2 \gamma  \sqrt{ Y(Y+\eta) })
    \right]
\end{equation}
where we have defined $\gamma \equiv \sqrt{ \frac{ \tas\,p}{2} }$.

When $Y$  is large and $\frac{\eta}{Y}$ is small,
we may rewrite this formula by first employing the 
asymptotic expansion of the Bessel function and subsequently 
expanding as a function of $\frac{\eta}{Y}$ 
\begin{equation}
\label{assymptotic}
   \tilde{R}_1(\eta, Y_s) \,= \, 
\frac{ 
       \tilde{r}^{(0)} p 
     } {\gamma} \,
     \frac{1}{\sqrt{\pi}}
     \frac{ e^{2 \gamma \,Y_s } } 
     {  (\gamma \,Y_s )^{3/2} } 
            e^{-\gamma\, \eta}\, \left[\,1 + \gamma\,\eta \, \right] \;,
\end{equation}
where  we have defined the rapidity at the 
saturation boundary 
\be
Y_s \equiv \ln \,\frac{z\,s}{Q_s^2} \;.
\ee
Re-expressing this result in terms of 
$z$ and $zs\,R(\un{x},z) = 
\tilde{R}(\un{x},z) = \theta\left(1-x^2_\perp Q_{s}^2\right) 
\tilde{R}_1(\un{x},z) $, we have
\begin{equation}
\label{Rzs}
  R(\un{x},z) \,=\, \frac{\tilde{r}^{(0)}}{\gamma^{5/2} \sqrt{\pi}} \frac{1}{z s} \left(\frac{z\,s}{Q_{s}^2}\right)^{2\gamma} \left(x_\perp^2 Q_{s}^2\right)^{\gamma} 
  \left[ \frac{1 - \gamma \log(x_\perp^2Q_s^2)} {
  \left(\log \frac{zs}{Q_{s}^2}\right)^{3/2} 
  } \right] \,\theta\left(1-x^2_\perp Q_{s}^2\right) \;.
\end{equation}

Next we can estimate the behavior of $F_{2}^{val}(x_{Bj}, Q^2)$ 
at large $Q^2$ and small $x_{Bj}$.  To determine $F_{2}^{val}(x_{Bj},Q^2)$ 
we  substitute  \eq{photonsplit}
for photon wave function  
and \eq{Rzs} for
$R(z,\un{r})$ 
into \eq{f2} for $F_{2}^{val}$. 
For large $Q^2$ the 
integrand falls off very rapidly due to asymptotic descent of 
the modified Bessel functions. Thus the integral is dominated by 
the region when $\epsilon^2 r^2 = Q^2r^2 z(1-z)$ is small. 
This occurs when $z < \frac{1}{r^2 Q^2}$  for $r$ larger than  
$\frac{1}{Q^2}$.
(The region near $z=1$ is suppressed 
relative to $z=0$ as can be seen from \eq{Rzs} .) 
For small values of $\epsilon r$ we approximate  
$K_1(\epsilon r)\approx \frac{1}{\epsilon r}$  and neglect
the corresponding contribution from 
$K_0(\epsilon r) \approx -\log(\epsilon r)$ to find
\begin{equation}
   F_{2}^{val}(x_{Bj}, Q^2) = \frac{Q^2}{4\pi s} 
 \int_{\frac{1}{Q^2}}^{\frac{1}{Q_{s}^2}}  \frac{d^2r}{r^2} 
 \int_{0}^{\frac{1}{r^2 Q^2}} \frac{dz}{z} \tilde{R}_1(\un{r}, z) \;.
\end{equation}
Next we neglect the logarithmic dependencies in the square
brackets of \eq{Rzs} and
perform the integrals over $z$ and $r$. The
result for $F_2(x,Q^2)$ is
\begin{equation}
   F_{2}^{val}(x,Q^2) \sim  \left( \frac{Q^2}{Q_s^2} \right)^{\gamma} \,
                x \left(\frac{1}{x}\right)^{2\gamma}
\end{equation}
Using the leading twist relation between $F_{2}^{val}$ and the 
valence quark phase space distribution \cite{Jaffe}
\begin{equation}
     F_2^{val}(x, Q^2) = \int^{Q^2} dk^2\, x \frac{dn_{val}}{dx \,d^2k} \; ,
\end{equation} 
we may differentiate $F_2^{val}$ and determine the parametric 
behavior of the unintegrated distribution function.
\begin{equation}
\label{phasespacedist}
   \frac{dn_{val}}{dx \, d^2k} \sim \left(\frac{Q_s^2}{k^2}\right)^{1-\gamma}  
                 \left(\frac{1}{x}\right)^{2\gamma} \;.
\end{equation}

A number of features warrant discussion. First, quantum evolution
generates an anomalous dimension of $-\gamma$ which enhances the
valence quark distribution by a factor of
$\sim\left(\frac{Q_s^2}{k^2}\right)^{-\gamma}$ relative to the naive
distribution $\sim\left(\frac{Q_s^2}{k^2}\right)$. Second, the
intercept found for the linear case in Sect.~\ref{linear} , $\alpha_R
= \sqrt{2 \tas}$, differs from the intercept found here where
saturation effects were fully included, 
\be\label{satint}
\alpha_R = \sqrt{\frac{2 \tas}{1-\kappa}}.
\ee  
The factor $(1-\kappa)^{-\frac{1}{2}}$, reflects the fact that the
saturation scale $Q_{s}^2 = \Lambda^2 e^{\kappa
\tau}$ is serving as an infrared cutoff which increases with collision
energy.

\section{Relation to RHIC data on baryon stopping}
\label{dataSection}

The discussion presented above addressed the issue of valence quark
distributions in a single nucleus. To compare obtained results with
the experimental data produced in heavy ion collisions at RHIC one has
to calculate valence quark production cross section. In principle the
problem can be formulated in a way similar to the gluon production
problem in the saturation framework \cite{claa,KM,kv,yuaa}. First one
should solve the quasi-classical problem of including all multiple
rescatterings in the valence quark production cross section
\cite{claa,KM,kv,yuaa} and then one should continue by including the 
effects of quantum evolution in the obtained expression \cite{KT}. The
above program has been carried out for gluon production in DIS and pA
collisions in \cite{KM,KT}. However for nuclear (AA) collisions the
gluon production problem complicates tremendously and still remains to
be solved \cite{kv,yuaa}.  Here we are not going to try to solve the
problem of valence quark production. Rather we are going to make some
$qualitative$ comparisons with the $AA$ net-proton data assuming that
the produced net baryons are simply proportional to the sum of the
valence quarks in the incoming nuclear wave functions.

The two nuclei collide with beam rapidities $Y_B$ and $-Y_B$. At
rapidity $y$, the valence quark content (per unit rapidity) of the
right moving wave function is $x_{R} f_{val}(x_{R}, Q_{s}^2)$, where
$x_{R}=e^{-(Y_B-y)}$. Similarly the valence quark content of the left
moving wavefunction is $x_{L} f_{val}(x_{L}, Q_{s}^2) $ where $x_{L} =
e^{-(Y_B+y)}$. Thus we expect the net baryon number to scale as
\begin{equation}
    \frac{dN^{net}_B}{dk^2 dy} \sim  x_{R} f_{val}(x_{R},Q_{s}^2) + x_{L} f_{val} (x_{L}, Q_{s}^2) 
\end{equation}
We will parametrize, $x f_{val}(x,Q_{s}^2)$ with a power law   
$\sim \left(x\right)^{\Delta_R}$ . This  parametrization is motivated
by Regge phenomenology and $\Delta_R$ is therefore referred to as the 
Reggeon intercept below.  
Therefore the 
rapidity dependence of the net baryons is given by 
\begin{equation}
\label{produce}
    \frac{dN^{net}_B}{dk^2 dy} \sim  e^{-\Delta_R (Y_B-y)} + 
e^{-\Delta_R (Y_B+y)} \; .
\end{equation}
In the small $x$ regime we expect the  Reggeon intercept 
to be given by the formula 
\begin{equation}
\label{gamma}
\Delta_R =  1- 2\gamma \equiv  1 - 2\,\sqrt{\frac{\alpha_s C_F}{2\pi}}\;.
\end{equation}

\begin{figure}
\begin{center}
\epsfxsize=10cm
\leavevmode
\hbox{ \epsffile{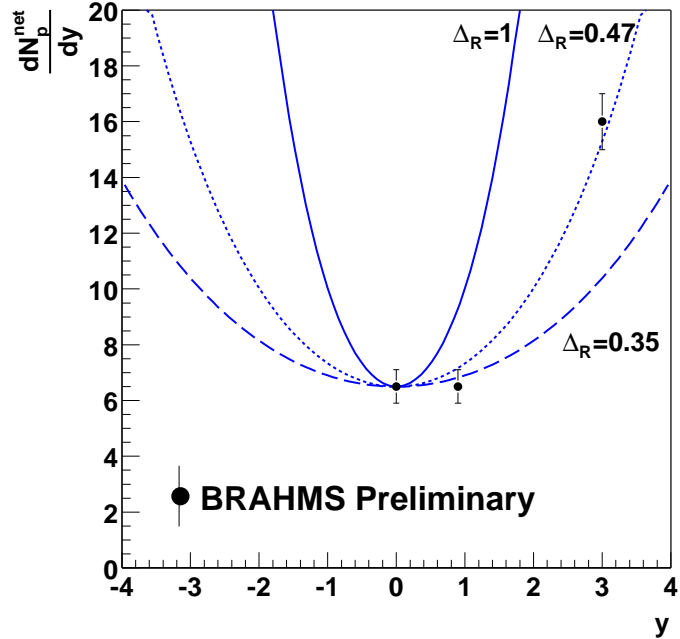}}
\end{center}
\caption{
A comparison to preliminary BRAHMS data  
\protect\cite{BRAHMS} 
on net-proton rapidity distributions with the 
functional form given by \eq{produce}. 
}
\label{data}
\end{figure}

We have fitted the observed net baryon distribution with the functional 
form given by \eq{produce} and determined the Reggeon intercept.
Fig.~\ref{data} shows the fit curve with  
$\Delta_R=0.47$ together with the naive intercept $\Delta_R=1$. 
The naive intercept completely fails to reproduce the data. 
The remaining curve $\Delta_R=0.35$ will be discussed shortly.

The value of $\Delta_R = 0.47$ should be compared to the the
non-perturbative reggeon intercept. Indeed, within the context of
Regge theory valence quark transfer at high energy is given by the
$I=1$ Regge trajectory, $\rho, f_2, \omega_3,\dots\; $. Fits to the
$\pi^{-}p$ and $\pi^{+}p$ cross sections give \cite{DL}
\begin{eqnarray}
      \sigma_{tot}^{\pi^{-}p} &=& 13.63\, s^{0.0808} + 36.02\, s^{-0.4525} \\
      \sigma_{tot}^{\pi^{+}p} &=& 13.63\, s^{0.0808} + 27.56\, s^{-0.4525}
\end{eqnarray}
The difference in these cross sections is proportional to the forward
amplitude with valence quark exchange.  Thus in Regge theory the
exchange amplitude scales as, $R \sim s^{-0.4525} \sim x^{0.4525}$.
This prediction of Regge theory  
that $\Delta_R = 0.4525$ seems to
agree rather well with the BRAHMS preliminary net-proton rapidity
distributions.

We wish to compare the fitted intercept with  \eq{gamma}. The current state of
theory is  not sufficient to compare in detail with RHIC data on net-baryon
production.  In addition, many of the approximations of the Color Glass
Condensate are stretched in the kinematic window of the data.  Nevertheless, it
is  useful to compare \eq{produce} to the net-baryons produced in the RHIC
experiment in order to verify that this theoretical result is  not in immediate
contradiction with data.  

Employing the phenomenological analysis of \cite{KLN},
we estimate the saturation scale at the rapidity $y$ 
\begin{equation}
         Q_{s}^2(y) = Q_{s}^2 (0) e^{-\kappa  y }
\end{equation}
where $Q_{s}^2(0) = 2.05$~GeV$^2$  and $\kappa\approx 0.25$.
Thus at $y=3$ we find $Q_s^2(y=3)\approx 1.0$~GeV$^2$ which is
not a very large scale.  Very roughly then, $\alpha_s(Q_s)$ varies from  
$0.33-0.50$ as $y$ varies from $0-3$ units of rapidity  and is 
not particularly small.  Of course the calculations leading 
to \eq{gamma} require $\alpha_s \ll 1$ or $Q_{s}^2/\Lambda^2_{QCD} \gg 1$.  

Further we estimate $x$ in the kinematic window of the BRAHMS
data. For the right moving nucleus we have roughly, $x_{R} =
e^{-(Y_B-y)}$ which varies between $0.005-0.1$ in as $y$ varies
between $0-3$ units of rapidity. Thus $x_{R}$ is also not particularly
small as we move forward in rapidity.

Pressing onward, we substitute $\alpha_s = 0.5$ and $\alpha_s = 0.33$
into \eq{gamma} and determine $\Delta_R \approx 0.35 $ and $ \Delta_R
\approx 0.47$, respectively. Thus, provided the coupling is small, the  
perturbative Reggeon (\eq{gamma}) also reproduces the experimental
intercept although with many more qualifications than the
non-perturbative Reggeon.

Recent data from the $dAu$ run from the RHIC collider probed the
effects of gluon saturation in a nucleus
\cite{dAtaphob,dAtaphen,dAtastar,dAtabrahms}.  
The experiments show little or no
suppression of moderate to high $p_T$ hadrons at mid-rapidity
\cite{dAtaphob,dAtaphen,dAtastar,dAtabrahms}. These data do not favor the
prediction of high-$p_T$ hadron suppression made in
\cite{KLM} based on extending small-$x$ evolution to high-$p_T\sim Q_s$. 
Thus it would seem that the experiments have already ruled out
$\log(1/x)$ evolution in this kinematic domain. However, a number of
points should be considered.  First, baryon number evolution is
governed by double logarithms $\alpha_s \log^2(1/x)$ as opposed to
single logarithms $\alpha_s \log(1/x)$ in the gluon case. Thus it is
reasonable to hope that evolution effects are stronger in baryons than
the corresponding effects in gluons. Second, multiple rescatterings
\cite{KNST,BKW,JNV,AG,ktbroadening1,ktbroadening2,Vitev03,ktbroadening3}
which were not considered in \cite{KLM} introduce the Cronin enhancement
\cite{Cronin} in $R^{dA}$. The effect of small-$x$ evolution is to reduce 
this enhancement, eventually wiping out the Cronin effect at very high
energies \cite{AAKSW,KKT}.  However, the effect of small-$x$ evolution
at moderately high energy is to somewhat reduce the Cronin maximum
without eliminating it completely. Strictly speaking, the prediction
of high-$p_{T}$ hadron suppression is only for $p_{T} \gg Q_{s}$ and
beyond the Cronin maximum. For RHIC kinematics this means $p_T \agt
5.0$\,GeV which corresponds to rather large Bjorken $x\agt 0.06$.
This value of $x$ is significantly larger than the values of $x$
relevant to net-baryon production at mid-rapidity
$x\approx0.01$. Therefore, it is not obvious that $dAu$ data
constrains the bulk properties of net-baryon production calculated
here.  Away from mid-rapidity (at say $y=3$) $x$ becomes larger
$\approx0.1$, our calculation of net-baryon production is no longer
reliable, and indeed the experiments rule out strong evolution effects
for these values of $x$.  Between $y=0$ and $y=3$ the calculation
provides a qualitative guide to the relative contributions of
kinematic effects (leading to $\Delta_R=1$) and kinematic+quantum
effects (leading to $\Delta_R= 1-\sqrt{2 C_F \alpha_s/\pi}$).
For these reasons we feel that the comparison with data is instructive
if not completely justified.

\section{Conclusions}

We have studied how isospin and baryon number are transported to small
$x$.  In particular we have studied how parton saturation affects the
valence quark distribution. We first constructed the analog of the
McLerran-Venugopalan (MV) model for valence quarks
(Sect.~\ref{MVQUARK}). The model illustrates how multiple
rescatterings regulate the infrared singularities in the valence quark
distribution.  The saturation scale serves as an energy dependent
infrared regulator as for the gluon case.  For large transverse
momentum, the valence quark phase space distribution at the
quasi-classical level is
\begin{equation}\label{conc1}
       \frac{dn_{val}}{dy\, d^2k} \propto  x \, 
\left(\frac{Q_{s}^2}{k^2} \right)\; ,
\end{equation}
where $y = \log\left(\frac{1}{x}\right)$.

Employing Mueller's dipole framework we subsequently constructed a
small $x$ evolution equation for the forward scattering amplitude with
valence quark exchange between the dipole and the target
(Sect.~\ref{QUANTUM}).  This equation illustrates how saturation
influences the evolution of valence quark quantum numbers to small
$x$.  Indeed, as indicated by \eq{eqr1}, quantum evolution stops as
unitarity constraints set in.

Next we investigated the solutions of the small $x$ evolution equation
in the linear and non-linear regions (Sect.~\ref{linear} and
Sect.~\ref{nonlinearSection}).  In the linear region, the solution
reproduces the $x$ intercept found previously by summing ladder
diagrams with quark exchange\cite{KL}.  Quantum evolution enhances the
valence quark rapidity distribution relative to the MV model, changing
the $x$ dependence from $x$ to
\begin{equation}
\label{eqsum}
\left(x\right)^{1-\sqrt{\frac{2\as C_F}{\pi} }}. 
\end{equation}

Using a simple theta function model for the dipole scattering
amplitude $N(\un{x},\un{b},\tau)$, we then studied how parton
saturation and the dynamics of the Color Glass Condensate influence
the $x$ evolution of valence quarks. As in the MV model, we found that
the saturation scale acts as an infrared regulator which increases
with energy as $Q_s^2 = \Lambda^2 e^{\kappa \tau}$. The effect of
quantum evolution is to change the canonical dimensions of the valence
quark distribution.  For large transverse momentum, we found
\begin{equation}\label{conc2}
 \frac{dn_{val}}{dy\, d^2k} \propto 
 \left( \frac{Q_s^2}{k^2}\right)^{1-\gamma}x^{1 - 2\gamma},\;\; 
\mbox{where} \;\; \gamma  \equiv 
\sqrt{ \frac{\as C_F}{2\pi \,\left(1-\kappa\right)}  } \; .
\end{equation}
The intercept $\Delta_R = 1-2\gamma$ is very similar to the intercept
in the linear regime differing only by a factor of $(1-\kappa)^{-1/2}$
in $\gamma$. This difference reflects the increase of the saturation
scale with energy.

Finally we studied net-baryon rapidity distributions at RHIC and
extracted a phenomenological intercept from the data, $\Delta_R
\approx 0.47$.  This value is in line with the expectations of Regge
theory \cite{DL}.  For $\alpha_s\approx
\frac{1}{3}$ this value is also in agreement with the intercept of  \eq{eqsum}.
Whether the perturbative reggeon can quantitatively explain the
net-baryon data for phenomenologically relevant $x$ remains an open
question.  Exciting new data on valence quarks at small $x$ is coming
from the $d-Au$ run at the RHIC collider and this data will provide
new constraints which will ultimately settle this question.

\section*{Acknowledgments}

The authors would like to thank Andrei Belitsky,  Yuri
Dokshitzer, Jamal Jalilian-Marian, Alfred Mueller, Dam Son, and
Kirill Tuchin for helpful discussions.

The work of Yu. K. was supported in part by the U.S. Department of
Energy under Grant No. DE-FG03-97ER41014 and by the BSF grant $\#$
9800276 with Israeli Science Foundation, founded by the Israeli
Academy of Science and Humanities. The work of the remaining authors
was supported by the U.S. Department of Energy under Grant
No. AC02-94CH10886.

\appendix
\section{The Light Cone Wave Function}
\label{wavefunction}

In this appendix we calculate the valence quark distribution in light
cone wave function of a quark moving in the ``plus" direction. The
lowest order graph is given by Fig.~\ref{lo} . Following the rules of
light cone perturbation theory this graph is given by
\begin{equation}
\label{lolong}
    \psi^a_{ \sigma\lambda} (\un{k},\un{p}-\un{k},z)=
    \frac{\left\langle f | H_I |i \right\rangle/\sqrt{k^+p^+}} {
    \sum_{f} p^-_f -\sum_{i} p^-_i }
    = \frac{ g T^a 
       }{ (p-k)^- + k^- - p^- }
       \frac{ \bar{u}_{\sigma}(k) }{\sqrt{k^+}} 
       \gamma
       \cdot
       \epsilon^{\lambda}(p-k)  
       \frac{u_{\sigma}(p)}{\sqrt{p^+}} \;.
\end{equation}
Here $k^+ = z p^+$, $\sigma =\pm$ is the helicity of the quark,
$p^{-}=p^{0} - p^{z} = \frac{p_T^2}{p^+}$, and we have divided the
matrix element by $\sqrt{p^{+}k^{+}}$ as is conventional in the
definition of $\psi^a_{\sigma\lambda}$. We are working in the light
cone gauge $A^{+}$=0 where $\epsilon^\lambda(k) =
(\epsilon^+,\epsilon^-, \un{\epsilon}^\lambda) = (0, \frac{2
\un{k}\cdot\un{\epsilon}^{\lambda}}{k^+}, \un{\epsilon}^{\lambda} )$
and $\un{\epsilon}^{\lambda\pm} = (\frac{1}{\sqrt{2}}, \pm
\frac{i}{\sqrt{2}})$.  Using the formulas for the matrix elements of
spinors with definite helicities given in Table II of Ref.~\cite{lb}
\begin{eqnarray}
   \frac{\bar{u}_\sigma(p)}{\sqrt{p^+}} \gamma^{+} \frac{u_\sigma(q)}{\sqrt{q^+}}
   &=& 2 \\
   \frac{\bar{u}_\sigma(p)}{\sqrt{p^+}} \gamma^{i}_\perp \frac{u_\sigma(q)}{\sqrt{q^+}}
   &=& \frac{p_\perp^i - \sigma\, i \epsilon^{ij} p_\perp^j}{p^+} + \frac{q_\perp^i +  \sigma\, i \epsilon^{ij}q_\perp^{j}}{q^+} \; ,
\end{eqnarray}
a simple calculation reduces \eq{lolong} to the wave 
function given in the text
\be
\psi^a_{\sigma\lambda} ({\un k}, {\un p}- {\un k}, z) \, = \, g \, T^a \, 
[1 + z - \sigma \lambda (1 - z)] \
\frac{{\un \epsilon}^\lambda \cdot ({\un k} - z \, {\un p})}
{({\un k} - z \, {\un p})^2} \;.
\ee
 
\section{Initial Conditions for Quantum Evolution} 
\label{qbarq}

This appendix details the steps leading to \eq{qq} for the initial
conditions of the quantum evolution. 
\begin{figure}
\begin{center} 
\epsfxsize=5.0cm
\leavevmode
\hbox{ \epsffile{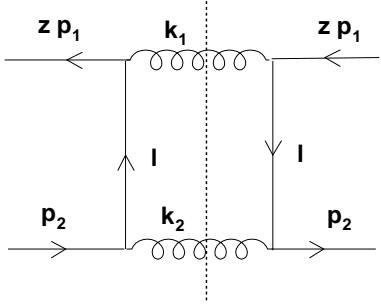}}
\end{center}
\caption{A forward scattering amplitude with $\bar{q}$ exchange.}
\label{figqq}
\end{figure}
First calculate the squared amplitude shown in Fig.~\ref{figqq},
summing over gluon helicities and colors, and averaging over the
quark and antiquark color and helicities
\begin{equation}
 \frac{1}{(2 N_c)^2} \sum \left|M\right|^2 = 
 \frac{C_{F}^2 g^4}{N_c} \frac{2 z s}{\un{l}^2} .
\end{equation}
Here we have assumed Regge kinematics (see e.g. \cite{rossbook}) where
$-l^2/s\ll1$ and where we may replace $l^2$ with $-\un{l}^2$.  Now
integrate over the loop momentum associated with the final state phase
space
\begin{equation}
 \int \frac{d^{4}l}{(2\pi)^4}\, 2\pi \delta_+(k_1^2) \,
2\pi\delta_+(k_2^2) \; ,
\end{equation}
using the delta functions to eliminate the $l^+$ and $l^{-}$ integrals.
Dividing by  the flux factor $z p_1^{+}p_2^{-} = 2\,z s$ we 
find the total cross section with flavor exchange   
\begin{equation}
   \sigma_{val}^{q \bar{q}} =  \frac{2\, C^2_{F}\as^2}{N_c \,zs} \int 
\frac{d^2\un{l}}{\un{l}^2} \; .
\end{equation}
The upper and lower limits of integral over $\un{l}$ are discussed in
the text.  To relate $\sigma^{q \bar{q}}_{val}$ to a cross section of
a dipole on a nucleus at a given impact parameter
$\frac{d^2\sigma^{q\bar{q}-A}_{val}}{d^2b}$ , we must multiply this
cross section by the flux of valence quarks in the incoming nucleus,
$(N_c\,A)/S_\perp$.  Thus we find
\begin{equation}
   2 \,R(\un{x},\un{b}, z) = \frac{d^2\sigma_{val}^{q\bar{q}-A} }{d^2b}  
                   =   \frac{A} {S_\perp} \frac{2\, C_{F}^2\as^2}{zs}  \int \frac{d^2l}{l^2} \; .
\end{equation}

\section{Solution of the model non-linear evolution equation} 
\label{solution}

Our goal in this appendix is to solve the model non-linear evolution
equation which replaces the scattering amplitude with a theta
function. The model is described fully in
Section~\ref{nonlinearSection} where an evolution equation for
${\tilde R}_1(\eta,Y)$ was obtained
\be\label{oreqapp}
{\tilde R}_1 (\eta, Y) \, = \, {\tilde r}^{(0)} \, p \, (Y+\eta) +
\frac{\tas \, p}{2} \, \int_0^{Y} d Y' \,
\int_0^{\eta + Y - Y'} d \eta' \, {\tilde R}_1 (\eta', Y') \;.
\ee
Here we have defined  the following variables 
which are described fully in the text:
$\tau \,  \equiv  \, \ln \frac{z \, s}{\Lambda^2},  \;
Q_s^2 \, \equiv \, \Lambda^2 \, e^{\kappa \tau}, \;
\eta   \equiv   \ln \frac{1}{x_\perp^2 Q_s^2}, \; 
Y \, \equiv \, \ln (z \, s \, x_\perp^2 ), $ 
 and $p \, \equiv \frac{1}{1-\kappa} \;.$
$\kappa$ is the intercept and is usually 
taken to be $0.2 - 0.3$.  From these definitions 
we derive: 
$\frac{\tau}{p} = \ln \frac{z s}{Q_s^2},$ and $Y = \frac{\tau}{p} - \eta \;.$ 
To solve this integral equation we  reduce it to 
a partial differential equation, deduce the  boundary conditions
using the original integral equation, and finally solve 
the differential equation employing Laplace transforms.

Differentiating \eq{oreqapp} with respect to
$\eta$ we get
\be\label{etaeq}
\frac{\partial
{\tilde R}_1 (\eta, Y)}{\partial \eta} \, = \, {\tilde r}^{(0)} \, p +
\frac{\tas \, p}{2} \, \int_0^Y d Y' 
\, {\tilde R}_1 (\eta + Y - Y', Y').
\ee
Differentiating \eq{oreqapp} with respect to $Y$ we get
\be\label{Yeq}
\frac{\partial
{\tilde R}_1 (\eta, Y)}{\partial Y} \, = \, {\tilde r}^{(0)} \, p + 
\frac{\tas \, p}{2} \, \int_0^\eta d \eta' \, 
{\tilde R}_1 (\eta', Y) + \frac{\tas \, p}{2} \,
\int_0^Y d Y' {\tilde R}_1 (\eta + Y - Y', Y').
\ee
Subtracting \eq{etaeq} from \eq{Yeq} we end up with
\be\label{int}
\frac{\partial
{\tilde R}_1 (\eta, Y)}{\partial Y} \, = \, \frac{\partial
{\tilde R}_1 (\eta, Y)}{\partial \eta} + \frac{\tas \, p}{2} \,
\int_0^\eta d \eta' {\tilde R}_1 (\eta', Y).
\ee
Differentiating \eq{int} with respect to $\eta$ we get
\be\label{pde}
\frac{\partial^2
{\tilde R}_1 (\eta, Y)}{\partial Y \, \partial \eta} \, = \, \frac{\partial^2
{\tilde R}_1 (\eta, Y)}{\partial \eta^2} + \frac{\tas \, p}{2}{\tilde R}_1 
(\eta, Y).
\ee
The initial conditions for \eq{pde} are given by
\be\label{init1}
{\tilde R}_1 (\eta, Y = 0) \, = \, {\tilde r}^{(0)} \, p \, \eta, 
\ee
which follows from \eq{oreqapp},
\be\label{init2}
\frac{\partial {\tilde R}_1 (\eta, Y)}{\partial Y} \Big |_{Y = 0} 
\, = \,  {\tilde r}^{(0)} \, p \, \left( 1 + \frac{\tas \, p \, \eta^2}{4} 
\right), 
\ee
which follows from \eq{Yeq} combined with \eq{init1}, and
\be\label{init3}
\frac{\partial {\tilde R}_1 (\eta, Y)}{\partial Y} \Big |_{\eta = 0} \, 
= \, \frac{\partial {\tilde R}_1 (\eta, Y)}{\partial \eta} \Big |_{\eta = 0},
\ee
which follows from \eq{int}. The solution of \eq{pde} is uniquely
specified by three boundary conditions [Eqs. (\ref{init1}),
(\ref{init2}) and (\ref{init3})], which one can see explicitly by
putting it on the lattice in $\eta, Y$ space.

Let us search for solution of \eq{pde} in the following form
\be\label{guess}
{\tilde R}_1 (\eta, Y) \, = \, \int \frac{d \lambda}{2 \pi i} \,
e^{\omega(\lambda) Y + \lambda \eta} \, {\tilde R}_{\lambda},
\ee
where $\omega(\lambda)$ is some unknown function of $\lambda$ and the
$\lambda$-integral runs parallel to the imaginary axis to the right of
the origin. Plugging \eq{guess} into \eq{pde} we easily obtain
\be
\omega(\lambda) \, = \, \lambda + \frac{\tas \, p}{2 \lambda}
\ee
so that \eq{guess} becomes
\be\label{solut1}
{\tilde R}_1 (\eta, Y) \, = \, \int \frac{d \lambda}{2 \pi i} \,
\exp \left[\left( \lambda + \frac{\tas \, p}{2 \lambda} \right) Y 
+ \lambda \, \eta \right] \, {\tilde R}_\lambda,
\ee
where ${\tilde R}_\lambda$ should be fixed by initial conditions
[Eqs. (\ref{init1}), (\ref{init2}) and (\ref{init3})]. They translate
into three equations, which are correspondingly
\be\label{in1}
\int \frac{d \lambda}{2 \pi i} \, e^{\lambda \eta} \,  {\tilde R}_\lambda 
\, = \, {\tilde r}^{(0)} \, p \, \eta,
\ee
\be\label{in2}
\int \frac{d \lambda}{2 \pi i} \, e^{\lambda \eta} \,  
\left( \lambda + \frac{\tas p}{2 \lambda} \right) \, {\tilde R}_\lambda 
\, = \, {\tilde r}^{(0)} \, p \, \left( 1 + \frac{\tas \, p \, \eta^2}{4} 
\right),
\ee
and
\be\label{in3}
\int \frac{d \lambda}{2 \pi i} \, \exp \left[\left( \lambda 
+ \frac{\tas \, p}{2 \lambda} \right) Y \right] \, \frac{1}{\lambda} 
\, {\tilde R}_\lambda \, = \, 0.
\ee
Eqs. (\ref{in1}) and (\ref{in2}) fix ${\tilde R}_\lambda$ to be
\be\label{poly}
{\tilde R}_\lambda \, = \, \frac{{\tilde r}^{(0)} \, p}{\lambda^2} + 
\sum_{n=1}^\infty \, a_n \, \lambda^n,
\ee
where $a_n$'s are arbitrary constants, i.e. ${\tilde R}_\lambda$ is
fixed by the first two conditions up to an analytic function, which is
specified only by the third condition [\eq{in3}]. Next, substitute 
\eq{poly} into \eq{in3} and perform the contour integrals  
using the integral representation 
of the modified Bessel function
\footnote{This relation may be verified by expanding the exponential in $z$
and performing the integral term by term to obtain the Taylor series of 
$I_\nu(z)$.}
\begin{eqnarray}
I_\nu(z) =  \frac{\left( \frac{1}{2} z \right)^\nu}{2\pi i} 
           \int_{-\infty}^{(0+)} 
            \, \lambda^{-1-\nu}
              \exp\left(\lambda + \frac{z^2}{4\,\lambda}\right) \,d\,\lambda \;.
\end{eqnarray}
where the contour starts at $-\infty$ above the real axis,  circles 
origin, and returns to $-\infty$ below the real axis. We obtain
\be\label{polyf}
0 \, = \, \frac{2 {\tilde r}^{(0)}}{\tas} \, I_2 (\sqrt{2 \tas \, p \,
Y^2}) + a_1 \sqrt{\frac{\tas \, p}{2}} \, I_1 (\sqrt{2 \tas \, p \,
Y^2}) + a_2
\frac{\tas \, p}{2} \, I_2 (\sqrt{2 \tas \, p \, Y^2}) + \mbox{higher
order }I_m's.  
\ee 
Since all of the  Bessel functions in \eq{polyf} are linearly independent and
depend upon $Y$ only,  the coefficient in front of each of
$I_\nu$ should be $0$. Enforcing this condition we end up with
\be\label{a2}
a_2 \, = \, - \frac{4 \, {\tilde r}^{(0)}}{p \, \tas^2}, \hspace*{1cm}
a_n \, = \, 0 \, \, \, \,  \mbox{for} \, \, \, \, n \neq 2. 
\ee
We thus have
\be
{\tilde R}_\lambda \, = \, \frac{{\tilde r}^{(0)} \, p}{\lambda^2} -  
\frac{4 \, {\tilde r}^{(0)}}{p \, \tas^2} \lambda^2,
\ee
which, together with \eq{solut1} gives
\be\label{solut2}
{\tilde R}_1 (\eta, Y) \, = \, \int \frac{d \lambda}{2 \pi i} \,
\exp \left[\left( \lambda + \frac{\tas \, p}{2 \lambda} \right) Y 
+ \lambda \, \eta \right] \, {\tilde r}^{(0)} \, p \, \left(
\frac{1}{\lambda^2} - \frac{4}{p^2 \, \tas^2} \lambda^2 \right). 
\ee
Integrating over $\lambda$ in \eq{solut2} yields the
solution quoted in the text 
\begin{equation}
{\tilde R}_1 (\eta, Y)\,=\, 
\frac{ 
          \tilde{r}^{(0)} p 
        } {\gamma} \,
    \left[ \frac{Y+\eta} 
                { \sqrt{Y(Y+\eta)}  } \,
                I_1(2 \gamma  \sqrt{Y(Y+\eta)})
               - 
           \frac{ (Y(Y+\eta))^{3/2} }
                       {(Y+\eta)^3}  \, 
                I_{3}(2 \gamma  \sqrt{ Y(Y+\eta) })
    \right]
\end{equation}
where we have defined $\gamma \equiv \sqrt{ \frac{ \tas\,p}{2} }$.

\end{document}